\documentclass[final,3p,times]{elsarticle}
\usepackage{amsfonts}
\usepackage{mathrsfs}
\usepackage{amsmath}
\usepackage{xcolor}
\usepackage{amssymb}
\usepackage{bm}
\usepackage{slashed}
\usepackage{graphicx}
\usepackage{subfigure}
\usepackage{verbatim} 
\usepackage[normalem]{ulem}
\usepackage[linktocpage=true,bookmarks=true,bookmarksnumbered=true,breaklinks=true,pdfpagemode=Fullscreen,pdfstartview=FitBH]{hyperref}

\numberwithin{equation}{section}
\allowdisplaybreaks[4]

\definecolor{gesfpurple}{rgb}{0.47,0.19,0.42}

\definecolor{gesflanse}{rgb}{0.00,0.50,0.50}

\definecolor{gesfblue}{rgb}{0.08,0.42,0.76}

\definecolor{gesfred}{rgb}{1,0,0}

\definecolor{gesfwhite}{rgb}{1,1,1}

\definecolor{gesfblack}{rgb}{0,0,0}

\newcommand{\gsec}[1]{{\hypersetup{linkcolor=red}Sec.\,\ref{#1}\hypersetup{linkcolor=blue}}}

\newcommand{\geqn}[1]{\hypersetup{linkcolor=blue}(\ref{#1})\hypersetup{linkcolor=blue}}
\newcommand{\gfig}[1]{{\hypersetup{linkcolor=violet}Fig.\,\ref{#1}\hypersetup{linkcolor=blue}}}
\newcommand{\gtab}[1]{{\hypersetup{linkcolor=gesflanse}Table\,\ref{#1}\hypersetup{linkcolor=blue}}}

\newcommand{\FR}[2]{\displaystyle\frac{\,{#1}\,}{#2}}
\newcommand{\fr}[2]{\mbox{$\frac{\,{#1}\,}{#2}$}}

\def\bge{\begin{equation}}
\def\ede{\end{equation}}
\def\bga{\begin{aligned}}
\def\eda{\end{aligned}}
\def\bgp{\begin{pmatrix}}
\def\edp{\end{pmatrix}}
\def\bgs{\begin{subequations}}
\def\eds{\end{subequations}}
\newcommand{\beq}{\begin{equation}}
\newcommand{\eeq}{\end{equation}}
\newcommand{\bq}{\begin{equation}}
\newcommand{\eq}{\end{equation}}
\newcommand{\ba}{\begin{array}}
\newcommand{\ea}{\end{array}}
\newcommand{\beqa}{\begin{eqnarray}}
\newcommand{\eeqa}{\end{eqnarray}}
\newcommand{\beqs}{\begin{subequations}}
\newcommand{\eeqs}{\end{subequations}}

\def\di{{\mathrm{d}}}

\def\ii{\texttt{i}}
\def\[{\left[}
\def\]{\right]}
\def\({\left(}
\def\){\right)}

\def\REE{\text{Re}}
\def\IMM{\text{Im}}

\def\over{\overline}

\def\leqq{\leqslant}
\def\geqq{\geqslant}

\def\al{\alpha}

\def\X{\chi}

\def\ep{\epsilon}
\def\lam{\lambda}

\def\SU{\text{SU}}
\def\U{\text{U}}
\def\TT{\mathbb{T}}

\def\S{\mathcal{S}}
\def\HT{\widetilde{H}}
\def\SR{S_0^{}\!}
\def\Sp{\mathcal{S}_+^{}}
\def\Sm{\mathcal{S}_-^{}}
\def\X{\chi}
\def\MX{M_{\chi}^{}}
\def\T{\mathcal{T}}
\def\Tp{\mathcal{T}'}

\def\fr{\frac}

\def\Zz{\mathbb{Z}_2^{}}

\def\Mp{M_{\text{Pl}}}
\def\gaga{\gamma\gamma}
\def\XXX{\mathbb{X}}
\def\yt{\tilde{y}}

\begin{document}

\begin{frontmatter}

\setcounter{footnote}{0}
\renewcommand{\thefootnote}{\fnsymbol{footnote}}

\title{{\bf Realizing Dark Matter and Higgs Inflation in Light of LHC Diphoton Excess}}

\author{
{\sc Shao-Feng Ge}\,\footnote{gesf02@gmail.com}$^{a}$,~~
{\sc Hong-Jian He}\,\footnote{hjhe@tsinghua.edu.cn}$^{b,c}$,~~
{\sc Jing Ren}\,\footnote{jren@physics.utoronto.ca}$^{d}$,~~
{\sc Zhong-Zhi Xianyu}\,\footnote{xianyu@cmsa.fas.harvard.edu}$^{e}$
\vspace*{3mm}
}

\address{
$^a$\,Max-Planck-Institut f\"{u}r Kernphysik, Heidelberg 69117, Germany
\\[1mm]
$^b$\,Institute of Modern Physics and Center for High Energy Physics,
Tsinghua University, Beijing 100084, China
\\[1mm]
$^c$\,Center for High Energy Physics, Peking University, Beijing 100871, China
\\[1mm]
$^d$\,Department of Physics, University of Toronto, Toronto, Ontario, Canada M5S1A7
\\[1mm]
$^e$\,Center of Mathematical Sciences and Applications, and Department of Physics,
\\
Harvard University, Cambridge, Massachusetts 02138, USA
}

\begin{abstract}
LHC Run-2 has provided intriguing di-photon signals of a new resonance around $750\,$GeV,
which, if not due to statistical fluctuations, must call for new physics beyond
the standard model (SM) at TeV scale.
We propose a minimal extension of the SM with a complex singlet scalar $\,\mathcal S\,$
and a doublet of vector-like quarks. The scalar sector respects CP symmetry,
with its CP-odd imaginary component $\,\X$\, providing
a natural dark matter (DM) candidate. The real component of $\,\mathcal S\,$
serves as the new resonance ($750\,$GeV) and explains the diphoton excess of the LHC Run-2.
The new scalar degrees of freedom of $\,\mathcal S\,$ help to stabilize the Higgs vacuum,
and can realize the Higgs inflation around GUT scale, consistent with the current
cosmological observations. We construct two representative samples
A and B of our model for demonstration.
We study the mono-jet signals of DM production from invisible decays
$\,\REE(\S)\to \X\X\,$ at the LHC Run-2. We further derive the DM relic density bound, and
analyze constraints from the direct and indirect DM detections.
\\[1.5mm]
\hfill Phys.~Lett.~B (2016) in Press [arXiv:1602.01801]
\end{abstract}

\end{frontmatter}

\graphicspath{{figs/}}

\renewcommand{\thefootnote}{\arabic{footnote}}
\setcounter{footnote}{0}

\vspace*{5mm}
\section{Introduction}
\label{sec:1}

Both ATLAS and CMS collaborations newly reported intriguing
di-photon excess around 750\,GeV in $pp$ collisions at the LHC Run-2 \cite{ATLAS}\cite{CMS}.
With 3.2\,fb$^{-1}$ integrated luminosity, ATLAS observed a signal excess at
$\,M_{\gaga}^{}=747$\,GeV in the di-photon invariant mass distribution with a $3.9\sigma$
local significance by assuming a wide resonance width of about 45\,GeV. For the narrow
resonance assumption, the local significance reduces to $3.6\sigma$.
At the same time, CMS collected 2.6\,fb$^{-1}$ data set and found a di-photon excess
at $\,M_{\gaga}^{}=760$\,GeV with $2.6\sigma$ local significance
under narrow width assumption \cite{CMS}.
When taking this resonance as a narrow-width (pseudo)scalar particle $\,\mathbb{X}\,$
produced from gluon fusion, the LHC Run-1\,(8\,TeV) and Run-2\,(13\,TeV) data could be
combined to yield a di-photon excess,\,
$\,\sigma[gg\to\XXX\to\gaga ]=(4.6\pm 1.2)\,$fb,\,
at $\,M_{\gaga}^{}\approx 750$\,GeV \cite{fit}.\,
Despite the 125\,GeV Higgs discovery at the LHC Run-1\,\cite{Higgs} which seems
to complete particle spectrum of the standard model (SM) so far, this new anomaly
around $\,M_{\gaga}^{}\approx 750$\,GeV
would point to indisputable evidence of new physics beyond the SM at TeV scale
(if not due to statistical fluctuations or systematical errors).
Even though the experimental evidence is not yet compelling
and more data are expected from the upcoming LHC runs after the spring 2016,
it is well-motivated to explore new physics interpretations and implications of
such an intriguing anomaly, which will be invaluable guidelines for further
experimental tests in this year.

Since only spin-0 or spin-2 particles could decay into di-photons\,\cite{LY},
a scalar particle $\,\mathbb{X}\,$ with mass $\sim\!750$\,GeV would be
the simplest interpretation of this new resonance. A spin-2 massive Kaluza-Klein
graviton will couple to all SM particles with the same strength,
and is thus uneasy to explain the absence of di-boson signals of $WW/ZZ$
except the di-photon excess in the current Run-2 data.
There are already many recent papers studying various possible
explanations with scalar resonance and related new physics \cite{fit}\cite{750}.
In this work, we motivate this new resonance by resolving
two existing difficulties of the SM: the vacuum instability and the absence of
dark matter (DM) candidate. The SM Higgs potential suffers vacuum instability
at scales above $\sim\!10^{11}$GeV \cite{instable,1505.04825}, and new physics is needed to
stabilize the vacuum and realize successful cosmic inflation in the early universe.
In particular, the most economical approach of inflation is the Higgs inflation \cite{HI}\cite{HI2},
where the inflaton is identified as the SM Higgs boson, and including proper new physics
is required \cite{HINP,HX}. It was shown before that a minimal extension \cite{HX}
can save the Higgs inflation by introducing only a real singlet scalar and a vector-like
quark at TeV scale. The other serious defect of the SM is its lack of DM candidate
to provide the required 28\% composition of our universe.
For this work, we will present a minimal construction of new physics to resolve
three things altogether: (i) consistent realization of Higgs inflation around GUT scale;
(ii) natural DM candidate to explain the observed DM relic abundance; (iii) a new scalar
state with mass $\sim\!750$\,GeV to induce the enhanced di-photon excess at
the LHC Run-2 \cite{ATLAS}\cite{CMS}.
For this purpose, our minimal extension includes a complex singlet scalar $\,\S\,$
and a doublet of vector-like quarks with electric charges $\(\frac{5}{3},\,\frac{2}{3}\)$.\,
The scalar sector respects CP symmetry,
with the SM-like light Higgs boson $h\,(125\text{GeV})$ acting as the inflaton in the
early universe. The CP-odd imaginary component $\,\IMM (\S)$\, provides
a stable DM candidate, while the real component $\,\REE (\S)\,$
serves as the new resonance ($750\,$GeV), which is produced by gluon fusion via
vector-quark triangle loops, with di-photon decays to give the observed LHC excess.
The new scalar degrees of freedom of $\,\S\,$ help to stabilize the Higgs vacuum, and thus
realize successful Higgs inflation around GUT scale, consistent with the current
cosmology observation.

This paper is organized as follows. In section\,\ref{sec:2}, we construct a minimal
extension with a complex singlet scalar $\,\S\,$ and
a doublet of vector-like quarks $(\Tp\!,\,\T)^T$ at TeV scale.
Then, in section\,\ref{sec:3} we study the decays and production of the CP-even component of $\,\S\,$,\,
and realize the observed LHC di-photon signals at $\,M_{\gaga}^{}\simeq 750$\,GeV\,.\,
For explicit demonstration, we will construct two  representative samples A and B.\,
Section\,\ref{sec:4} is devoted to analyzing vacuum stability of the new Higgs potential,
and realizing a consistent Higgs inflation.
Next, we systematically analyze the CP-odd component of $\,\S\,$ as the DM candidate
in section\,\ref{sec:5}, where we will realize the observed DM relic abundance
in Sec.\,\ref{sec:5.1},
and study the DM production at the LHC Run-2 (\gsec{sec:5.2}), the DM direct detection
(\gsec{sec:5.3}), and the DM indirect detection (\gsec{sec:5.4}).
Finally, we conclude in section\,\ref{sec:6}.  Appendix\,A provides the needed formulas
for the $\,\REE(\S)$\, partial decay widths, while Appendix\,B presents
the additional one-loop $\beta$ functions induced by the new scalar couplings
and new Yukawa coupling.

\section{Model Setup with Singlet Scalar and Vector-like Quarks}
\label{sec:model}
\label{sec:2}

In this section, we construct a minimal model by implementing
a complex scalar singlet $\,\S\,$ and a doublet of vector-like quarks
$(\Tp\!,\,\T)^T$ at TeV scale.
As mentioned in \gsec{sec:1}, this can nicely tie three new physics ingredients altogether:
the consistent realization of Higgs inflation with the SM-like Higgs boson
$h\,(125\text{GeV})$ acting as inflaton, a stable DM candidate $\,\IMM (\S)$,\,
and a new scalar $\,\REE (\S)$\, of mass $\,750\,$GeV.\,
Hence, the newly observed di-photon excess from a 750\,GeV resonance decays
at the LHC Run-2 can link our predictions to the on-going DM detections and
the probe of Higgs inflation in the early universe.

Our Higgs sector consists of the SM Higgs doublet $\,H\,$ and a complex scalar singlet $\,\S\,$,
defined as follows,
\beqa
\label{eq:HS}
  H \,=
\left\lgroup
\begin{matrix}
  \pi^+ \\[1.5mm]
  \frac{\,v + h_0 \!+ \ii \pi_0\,}{\sqrt 2}
\end{matrix}
\right\rgroup ,
\qquad
\S \,=  \frac{\,u + S_0 + \ii\hspace*{0.3mm}\chi\,}{\sqrt 2}
   \,\equiv\,  \S_+^{} + \ii \S_-^{} \,,
\eeqa
where $\,v\,$ and $\,u\,$ denote the corresponding vacuum expectation values (VEV) of
$\,H\,$ and $\,\S\,$.\, Both the real and imaginary components,
$\S_+^{}$ and $\S_-^{}$, can help to stabilize the Higgs potential.
Under the CP transformation, we have $\,\S\to \S^*\,$,\,
which means $\,(\Sp,\,\Sm)\to (\Sp,\,-\Sm)\,$.\,
Namely, $\,\Sp \,(\,\SR\,)\,$ is CP-even and $\,\Sm \,(\,\X\,)\,$ is CP-odd.
Our construction imposes CP symmetry on the Higgs potential
as well as the Yukawa interactions of singlet $\,\S\,$.\,
As shown in Table\,\ref{tab:1}, we will further impose a separate $\Zz$ symmetry,
under which the Higgs doublet $\,H\,$ is even, and the singlet
$\,(\Sp,\,\Sm)\to (-\Sp,\,\Sm)\,$,\, i.e., $\,\S\to -\S^*\,$.\,
Thus, the building blocks of constructing a gauge-invariant and CP$\,\otimes\,\Zz$ symmetric
Higgs potential contain $\,H^\dag H$,\, $(\S\!+\!\S^*)^2$,\, and $(\S\!-\!\S^*)^2$,\,
where the second and third combinations are proportional to $\,\S^2_+\,$ and $\,\S_-^2$,\, respectively.
Hence, we can write down the following gauge-invariant and CP$\,\otimes\,\Zz$ symmetric
Higgs potential for $(H,\,\S)$\,,
\begin{eqnarray}
  V(H, \mathcal S)
& \!\!=\!\! &
- \mu^2_1 H^\dagger H
- \mu^2_2 \mathcal S^2_+
+ \lambda_1^{} (H^\dagger H)^2
+ \lambda_2^{} \mathcal S^4_+
+ \lambda_3^{} \mathcal S^2_+ H^\dagger H
\nonumber
\\[0.5mm]
& &
+ \mu^2_3 \mathcal S^2_-
+ \lambda_4 \mathcal S^4_-
+ \lambda_5 \mathcal S^2_+ \mathcal S^2_-
+ \lambda_6 \mathcal S^2_- H^\dagger H \,,
\label{eq:V}
\end{eqnarray}
where all masses and couplings are real.
We see that in the basis $(\Sp,\,\Sm)$,
the first line of our Higgs potential \geqn{eq:V} corresponds to the original Higgs
potential in Ref.\,\cite{HX} with a real singlet scalar.
Since the CP-odd pseudoscalar $\,\Sm\,$ has a positive mass-term $+\mu^2_3$ in
the potential \geqn{eq:V}, so it ensures a vanishing VEV of $\,\Sm\,$
and keep the CP symmetry intact in the scalar sector.
The CP-odd pseudoscalar $\,\Sm\,$ will serve as a stable DM candidate,
as to be analyzed in \gsec{sec:5}.
Note that scalar VEVs $(v,\,u)$ in \eqref{eq:HS} do not affect CP invariance,
except spontaneously breaking $\Zz$.\,  Since the mass term of
vector-like heavy quarks [cf.\ Eq.\,\eqref{eq:LY}] will softly break $\Zz$ and
contribute to the Higgs potential at loop level, this model is free from
the domain wall problem.

To explain the observed 750\,GeV excess of diphoton signals
from $\,gg\to S_0^{}\to\gaga\,$,\, the simplest natural resolution is to couple it
with certain charged quarks. But, a scalar singlet $\S$ cannot have gauge-invariant
and renormalizable Yukawa interactions with the SM fermions. Furthermore,
the LHC Run-2 has not found $S_0^{}$ decays into the SM fermions so far.
Hence, it is natural to couple the singlet $\S$ to certain new heavy quarks.
In our construction, we introduce a pair of vector-like quarks
$\,\TT = (\T'\!,\,\T)^T$,\,
which compose a weak doublet under the SM gauge group $SU(2)_L^{}$.
(The doublet vector-like quarks were invoked in model-buildings before\,\cite{TFSS}
with the SM hyercharge $\,Y = {1}/{6}\,$.\, For the present model, we extend it
to have hyercharge $\,Y = {7}/{6}\,$.\, This assignment was also considered in \cite{7/6}.)
The vector-like quarks $\,(\T'\!,\,\T)\,$
will induce production and decays of the new scalar $S_0^{}$ via triangle loops.
We arrange the quantum number assignments of our model in \gtab{tab:1},
where we have imposed a $\,\Zz\,$ symmetry to restrict the additional Yukawa interactions
involving the vector-like quark doublet $\,\TT\,$ and/or singlet scalar $\,\S\,$.\,
We conjecture that the singlet $\S$ interactions always conserve CP\,,\, and
all interaction forces respect $\Zz$.\, So the $\Zz$ symmetry could be softly broken
only via the bare mass term of the vector-like quark doublet $\TT$\,.\,

\begin{table}[t]
\setlength{\tabcolsep}{1.7mm}
\begin{center}
\caption{Quantum number assignments for the Higgs doublet $H$, the singlet scalar $\S$,
         the vector-like quark doublet $\,\TT =(\T', \T)^T$,\, and the SM quarks,
         under the SM gauge group $\,\SU (3)_C^{}\otimes \SU (2)_L^{}\otimes \U (1)_Y^{}\,$ and
         the discrete $\mathbb{Z}_2^{}$.\,
          All other fields have the same assignments as in the SM.
          Here $\,j\,(=1,2)\,$ stands for the indices of first and second family quarks,
          with $\,Q_{jL}^{}\!=(u_j^{},\,d_j^{})_L^T\,$\, and $\,Q_{3L}^{}\!=(t,\,b)_L^T\,$.}
\vspace*{3mm}
 \begin{tabular}{c||ccccccc|ccc}
   \hline\hline
   ~Groups~ &  $Q_{jL}^{}$  &  $u_{jR}^{}$ &  $d_{jR}^{}$  & $Q_{3L}^{}$ &  $t_{R}^{}$  & $b_{R}^{}$
   & $H$ & $\TT_L^{}$  &  $\TT_R^{}$  & $(\Sp,\Sm)$
   \\ \hline
   &&&&&&&&&&
   \\[-3.5mm]
   SU(3)$_C^{}$  & \bf{3} & \bf{3} & \bf{3} & \bf{3} &  \bf{3} & \bf{3}
   & \bf{1} & \bf{3} & \bf{3} & \bf{1}
   \\
   &&&&&&&
   \\[-3.5mm]
   SU(2)$_L^{}$  & \bf{2} & \bf{1} &  \bf{1} & \bf{2} & \bf{1} & \bf{1} & \bf{2}
   & \bf{2} & \bf{2} & \bf{1}
   \\
   &&&&&&&&&&
   \\[-3.5mm]
   U(1)$_Y^{}$  & $\fr{1}{6}$ & $\fr{2}{3}$ & $-\fr{1}{3}$ & $\fr{1}{6}$
   & $\fr{2}{3}$ & $-\fr{1}{3}$ & $\fr{1}{2}$ & $\fr{7}{6}$ & $\fr{7}{6}$ & 0
   \\
   &&&&&&&&&&
   \\[-3.5mm]
   $\mathbb{Z}_2^{}$
   & $+$ & $+$ & $+$ & $-$ & $-$ & $-$ & $+$ & $+$ & $-$
   & $(-,+)$
   \\
   \hline\hline
 \end{tabular}
\label{tab:assignments}
\label{tab:1}
\end{center}
\end{table}

The observed sizable di-photon rate at $\,M_{\gaga}^{}\approx 750$\,GeV
and the absence of dijet excess in the same mass region so far suggests that the new
resonance $\,\SR\,$ should have enhanced decay rate into di-photons. This indicates
that the heavy quarks may have larger electric charges and thus enhanced couplings
with di-photons. For this, we introduce a weak doublet of vector-like quarks,
$\,\TT = (\T'\!,\,\T)^T$,\, with hypercharge $\,Y = \fr{7}{6}\,$ and thus
the electric charges $\,\(\fr{5}{3},\,\fr{2}{3}\)$,\,
where the heavy quark $\,\T\,$ shares the same electric charge with the SM up-type quarks.

According to the model construction in \gtab{tab:1}, we write down the relevant
Yukawa interactions including the Yukawa interaction
between the vector-like quark doublet $\,\TT\,$ and
singlet scalar $\,\S$\,,\, as well as the Yukawa interactions between $\,\TT\,$
and light SM up-type quarks,
\beqa
\mathcal L_{\TT u}^{}
\,=\,
-y^{}_{ij}\over{Q}_{iL}^{}\HT u_{jR}^{}
- \yt_{j}^{} \overline{\mathbb T}_{L}^{} H u_{jR}^{}
- \fr{1}{2} \yt_{S}^{} \Sp \overline{\TT} \TT
- \fr{1}{2} M_0^{}\overline{\TT} \TT + \mbox{h.c.},
\label{eq:LY}
\eeqa
where $\,\HT =i\tau_2^{}H^*$,\, and
$\,i,j=1,2\,$ stand for flavor indices of the first and second family fermions.
We see that $\,\TT\,$ does not mix with third family top quark due to $\Zz$ symmetry.
The Yukawa coupling $\,\yt_{S}^{}$ in Eq.\,\eqref{eq:LY} is real,
since the singlet $\S$ interactions conserve CP\,. Besides, all interactions respect
$\Zz$ symmetry, and the only possible soft breaking term of $\,\Zz\,$
is the bare mass term ($M_0^{}$) of vector-like quark doublet $\,\TT\,$.\,
Eq.\,\eqref{eq:LY} gives the following mixing mass matrix for
$\,u_j^{}$ ($j=1,2$)\, and $\,\T$\,,\,\footnote{We also note that the small quark mixings between
the light families and the third family can arise from
dimension-5 effective operators involving singlet scalar,\,
$\,(y_{it}^{}/\Lambda){\Sp}\overline{Q}_{iL}^{}\HT t_R^{}\,$ and
$\,(y_{ib}^{}/\Lambda){\Sp}\overline{Q}_{iL}^{}H b_R^{}\,$,
where $i=1,2$ and $\Lambda$ is the cutoff.
Such effective operators will induce the desired small CKM mixings.
They may result from integrating out a heavy Higgs doublet $H'$
which is $\,\Zz$ odd and can realize dimension-4 Yukawa terms between the light
families and the third family,
$\,y_{it}'\overline{Q}_{iL}^{}\HT' t_R^{}\,$ and
$\,y_{ib}'\overline{Q}_{iL}^{}H' b_R^{}\,$.\,
Adding this heavy Higgs doublet $H'$ will increase
the scalar degrees of freedom and make vacuum stability much easier, but does not
change the main physics picture. For the current purpose of accommodating the diphoton excess,
we focus on the minimal setup for simplicity.}
\beqa
\label{eq:mass-mix}
  M_{u_j\mathcal T}^{}
=
  \fr{1}{\sqrt{2}\,}
\left\lgroup
\begin{matrix}
 y_{11}^{} v & y_{12}^{} v & 0 \\[1mm]
 y_{21}^{} v & y_{22}^{} v & 0 \\[1mm]
 \yt_{1}^{} v & \yt_{2}^{} v &
 \yt_{S}^{} u+\!\!\sqrt{2}M_0^{}
\end{matrix}
\right\rgroup ,
\eeqa
where the $(3,3)$-component contains both the VEV contribution term
$\,\yt_{S}^{} u/\!\sqrt{2}\,$
[from the third term of Eq.\,\eqref{eq:LY}] and the bare mass term $M_0^{}$
[from the fourth term of Eq.\,\eqref{eq:LY}].
For our purpose, we consider the parameter space of
$\,\yt_{j}^{}v \ll \yt_S^{}u+\!\!\sqrt{2}M_0^{}$\,.\,
Taking the small non-diagonal couplings $\yt_{1,2}^{}$ being comparable, we estimate
the small mixing of $\,\T\,$ and $\,u_{j}^{}$\,,\,
$\,\theta_{Lj}^{} \approx y_{jj}^{} \yt_{j}^{}v^2 / (\yt_S^{}u+\!\!\sqrt{2}M_0^{})^2$\,
for the left-handed quarks,
and $\,\theta_{Rj}^{} \approx \yt_j^{} v/(\yt_S^{}u+\!\!\sqrt{2}M_0^{})$\,
for the right-handed quarks.
Thus, we have nearly degenerate heavy quarks,
\beqa
M_{\T'}^{}\approx M_{\T}^{}\approx \fr{1}{\sqrt{2}\,}\yt_S^{}u+M_0^{}\,.
\eeqa
The small mixing couplings $\yt_j^{}$ will induce
$\T$ and $\T'$ decays. The heavy quark $\T$ has two main decay channels,
$\,\T \to u_j^{} h\,$ and $\,\T \to d_j^{} W^+$,\,
while $\,\T'$ dominantly decays via $\,\T' \to u_j^{}W^+$.\,
We find that for channels
$\,\T \to d_j^{} W^+\!,u_j^{}Z$\, and $\,\T' \to u_j^{}W^+$,\,
the decay amplitudes are dominated by the final state with longitudinal polarization
$W_L^+$.\, Since $M_{\T}^{}\gg M_W^{}$,\,
we can apply equivalence theorem\,\cite{ET} to compute the corresponding
Goldstone amplitudes with $W^+_L$ replaced by $\pi^+$.\,
Thus, we estimate the leading decay width for each channel as follows,
%
\begin{eqnarray}
\Gamma [\T \!\to\! u_j^{} h]\approx\frac{\yt_{j}^2}{16\pi\,}M_{\T}^{},
\quad~
\Gamma [\T \!\to d_j^{} W^+\!,u_j^{}Z]\approx
\frac{\,y_{jj}^2\theta_{Rj}^2\,}{32\pi}M_{\T}^{},
\quad~
\Gamma [\T' \!\!\to\! u_j^{} W^+]\approx\frac{\,\yt_{j}^2\,}{\,32\pi\,}M_{\T'}.
\quad~
\end{eqnarray}
It is clear that $\,\T\to u_j^{} h\,$ is the dominant decay mode for $\,\T$.
For later analysis, we will consider the parameter range,
$\,10^{-5}\lesssim \yt_j^{}\lesssim 10^{-3}$.\,
This is sufficient to evade the flavor constraints involving the first two family quarks,
and the tiny mixing coupling $\yt_j^{}$ is negligible
in our later analysis of renormalization group running and collider studies.
Furthermore, this ensures that the lifetimes of
$\T$ and $\T'$ are much smaller than $10^{-13}$s.\, So they are short-lived
and will have prompt decays inside the detector \cite{Buchkremer:2012dn}.
The searches of heavy vector-like quarks via prompt decays put nontrivial constraints
on the new quark masses. The limits on their decays into a light quark are weaker than
that into top or bottom. For $\,\T'\!\to u_j^{}W^+$,\,
the limit is $\,M_{\T'}^{}\!\gtrsim 690\,$GeV,
while the decay channel $\,\T\!\to u_j^{}h\,$ is much less constrained \cite{heavyQ}.

\section{New Particle Decays and Production}
\label{sec:S}
\label{sec:3}

In the physical vacuum, the Higgs doublet $H$ and singlet $\S$ acquire nonzero VEVs,
as shown in \eqref{eq:HS}. This spontaneously breaks
$\text{SU(2)}_L^{} \otimes \text{U(1)}_Y^{}\otimes\Zz$ down to U(1)$^{}_{\text{em}}$,
while the CP symmetry is retained.
The CP-even states $(h_0^{},\,S_0^{})$ can mix with each other via
$\,h = c_\alpha^{} h_0^{} + s_\alpha^{} S_0^{}$ and
$\,S = c_\alpha^{} S_0^{}\! - s_\alpha^{} h_0^{}$,\, where
$(c_\alpha^{},\, s_\alpha^{}) \equiv (\cos\alpha,\, \sin\alpha)$.\,
The mixing angle $\,\alpha\,$ is determined by diagonalizing the mass matrix,
\beqa
  \mathbb M^2_N \,=
\left\lgroup
\begin{matrix}
  \,2 \lambda_1^{} v^2 & \lambda_3^{} vu\,
\\[2mm]
  \,\lambda_3^{} vu & 2\lambda_2^{}u^2\,
\end{matrix}
\right\rgroup \!,
~~~\Longrightarrow~~~
  (\mathbb{M}^2_N)_{\text{diag}}^{} \,=
\left\lgroup
\begin{matrix}
  \,M_h^2 & 0\,
\\[2mm]
  \,0 & M_S^2\,
\end{matrix}
\right\rgroup \!,
\label{eq:massH}
\eeqa
with
\beqa
\label{eq:alpha}
\tan2\alpha \,=\, \frac{\lambda_3^{}vu}{\,\lambda_1^{}v^2 - \lambda_2^{}u^2\,}\,.
\eeqa
Alternatively, we may resolve the 3 involved scalar couplings
$(\lambda_1^{},\,\lambda_2^{},\,\lambda_3^{})$
in terms of the measured mass-eigenvalues $(M_h^{},\,M_S^{})\simeq (125,\,750)$\,GeV,
the known light Higgs VEV $\,v\simeq 246\,$GeV,\, and the Higgs mixing angle
$\,\alpha\,$ (which is taken as an input parameter, but will be constrained by the LHC data).
Thus, we have,
\beqa
\label{eq:lambda123}
\lambda_1^{}=\frac{\,M_h^2c_\alpha^2\!+\!M_S^2s_\alpha^2\,}{2v^2},
\quad~~
\lambda_2^{}=\frac{\,M_S^2c_\alpha^2\!+\!M_h^2s_\alpha^2\,}{2u^2},
\quad~~
\lambda_3^{}=\frac{\,s_\alpha^{}c_\alpha^{}(M_h^2\!-\!M_S^2)\,}{uv}.
\eeqa
Although the singlet scalar $\,\S\,$ does not couple to the SM fermions and gauge
bosons, the mixing between the two CP-even components $\,S_0^{}$ and $\,h_0^{}\,$ will induce
these couplings suppressed by $\,\sin\alpha\,$.\,
\gtab{tab:2} summarizes the coupling ratios
$\,\xi_{hXY}^{}\,$ and $\,\xi_{SXY}^{}\,$  relative to the SM counterparts,
for the mass-eigenstates $\,h\,$ and $\,S$.\,

\begin{table}[b]
\setlength{\tabcolsep}{2.7mm}
\caption{Coupling ratios $\xi_{hXY}^{}$ and $\xi_{SXY}^{}$ of the Higgs bosons $h$ and $S$,
         relative to the SM counterparts, where $\,V=W,Z$\,,\,
         and the SM Yukawa coupling is $\,y_f^{}=m_f^{}/v\,$.}
\centering
\begin{tabular}{c||ccc}
\hline\hline
$XY$ & $f \bar{f}$ & $VV$ & $\TT\overline{\TT}$
\\
\hline\hline
$\xi_{hXY}^{}$ & $c_\alpha^{}$ & $c_\alpha^{}$ & $s_\alpha^{}(\yt_S^{}\!/y_t^{})$
\\
\hline
$\xi_{SXY}^{}$ & $-s_\alpha^{}~$ & $-s_\alpha^{}~$ & $c_\alpha^{}(\yt_S^{}\!/y_t^{})$
\\
\hline\hline
\end{tabular}
\label{tab:Hig-coup}
\label{tab:2}
\vspace*{1.5mm}
\end{table}

Inspecting the cubic scalar coupling of $\,Shh\,$ vertex and using
Eq.\,\eqref{eq:lambda123}, we derive its compact form as follows,
\beqa
\label{eq:Shh}
G_{Shh}^{} =\,
\frac{\,s_\alpha^{}c_\alpha^{}(u c_\alpha^{}\!\!-\! v s_\alpha^{})\,}{u v}
(M_S^2\!+\! 2M_h^2)\,.
\eeqa
Clearly, this coupling is suppressed by $\,s_\alpha^{}\,$ for small mixing angle $\,\alpha\,$.
Hence, the decay width of $\,S \!\to hh\,$ is proportional to $\,s_\alpha^2\,$ and
will become negligible as $\,\alpha\!\to 0$\,.\,

\begin{figure}[t]
\centering
\includegraphics[height=6cm,width=7.3cm]{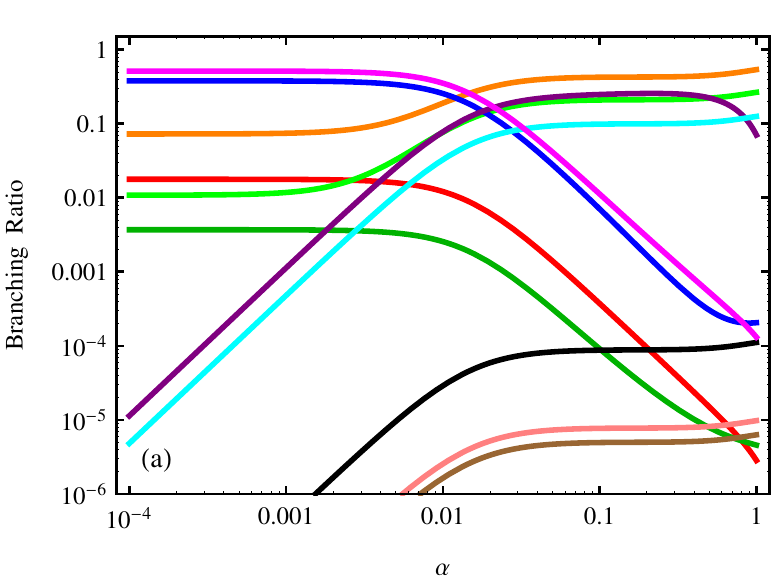}
\hspace*{-1mm}
\includegraphics[height=6cm,width=8.99cm]{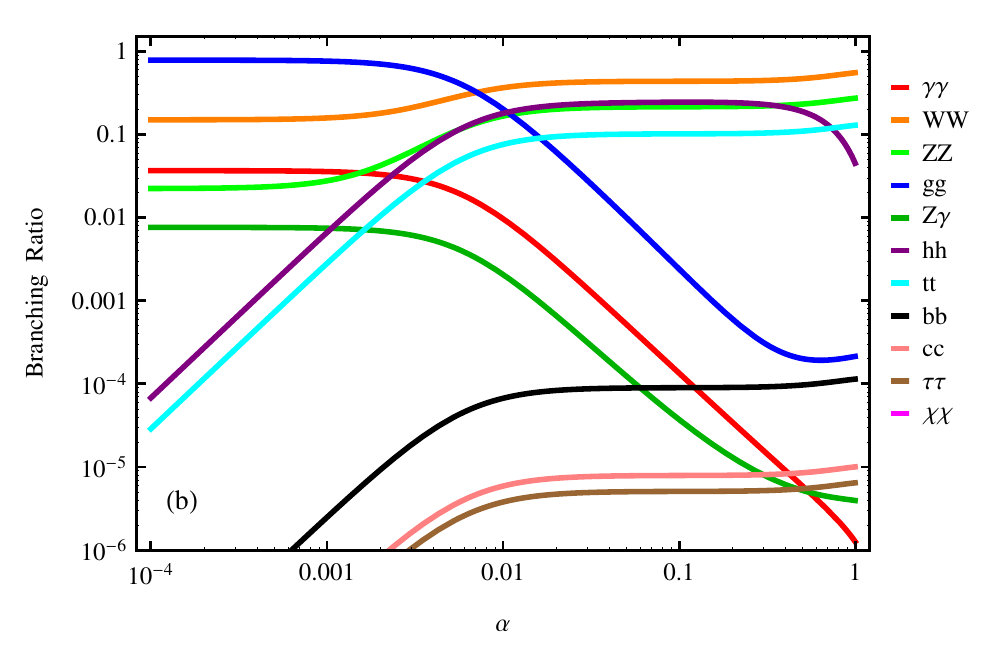}
\vspace*{-2mm}
\caption{Decay branching fractions of CP-even state $\,S\,(750\text{GeV})$\,
as a function of Higgs mixing angle $\,\alpha$\,.\,
We have input masses $\,M_{\T}^{} \simeq M_{\T'}^{}\! = 720$\,GeV for both plots,
as well as the VEV $\,u=0.57\,[1.2]\,$TeV and Yukawa coupling
$\,\yt_S^{} = 1.5\,[0.88]\,$ for plot-(a)\,[-(b)],
which are motivated by Sample-A\,[-B] of Eq.\,\eqref{eq:samples}.
In plot-(a) the invisible decay channel $\,S\to\X\X$\, is open under input
$\,\MX =240\,$GeV,\, while this channel is forbidden in plot-(b) due to
$\,\MX > M_S^{}/2\,$.\,
}
\label{fig:BR}
\label{fig:1}
\end{figure}

With \gtab{tab:2} and Eq.\,\eqref{eq:Shh},
we implement $S$ couplings into the decay width formula \geqn{eq:Gamma-scalar},
and compute the branching fractions of the CP-even state $S$ as a function of
Higgs mixing angle $\,\alpha\,$.\, Then, we present the $S$ decay branching
fractions in \gfig{fig:1}.
For a sizable mixing angle $\alpha$,\, we expect $S$ decays into
the SM fermions and gauge bosons to be significant.
\gfig{fig:1} shows that
for $\,\alpha \gtrsim 0.01$,\, the decay channels
$\,S\!\to\! WW,ZZ,hh\,$  and $\,S\!\to\! t\bar{t}\,$ dominate,
while $\,S\!\to\! \gaga\,$ and $\,S\!\to\! gg\,$  channels become much suppressed.
Hence, we see that in order to obtain a sizable branching fraction \,Br$[S\!\to\gaga]$\,
for enhanced diphoton rate at the LHC,
the Higgs mixing angle $\,\alpha\,$ should be fairly small, within the range of
$\,\alpha < 0.01\,$.\,

In the limit $\,\alpha\!\sim\! 0\,$ with the fixed VEV $\,u$\,,\,
we have, $\,\lambda_3^{}\!\sim\! 0\,$,\, $\,\lambda_1^{} \sim M^2_h/(2 v^2)\,$,
and $\,\lambda_2^{} \sim M^2_S /(2 u^2)$,\, according to Eq.\,(\ref{eq:lambda123}).
This implies that the two CP-even
states $\,h\,$ and $\,S\,$ nearly decouple from each other.
To summarize, imposing two extremal conditions on the Higgs potential $V$,
taking a small Higgs mixing angle $\,\alpha\,$,\, inputting the VEV
$\,v\,(\simeq 246\,$GeV),\,
and fixing Higgs masses $\,(M_h^{},\,M_S^{})\simeq (125,\,750)\,$GeV,
we have \,6\, conditions in total.
Thus, we can determine \,6\, parameters $(\mu^{}_1,\,\mu^{}_2,\,u)$\, and
$\,(\lambda_{1}^{},\,\lambda_{2}^{},\,\lambda_{3}^{})$\,
in the Higgs potential \eqref{eq:V}.
We are left with \,4\, free parameters
$(\mu^{}_3,\,\lambda_{4}^{},\,\lambda_{5}^{},\,\lambda_{6}^{})$\,
associated with masses and couplings of the CP-odd state \,$\X$\,.\,
We will use them to realize the viable Higgs inflation and DM in
\gsec{sec:4} and \gsec{sec:5}, respectively.

For demonstration, we construct two numerical samples
for our phenomenological study.
In both samples, we set inputs,
$\,M_S^{}\!=750\,$GeV,\, $\alpha \!=\! 10^{-3}$,\, and
\,$\theta_{L,R}^{}\!\simeq 0$\,.\,
The rest of parameters are defined as follows,
\begin{subequations}
\begin{eqnarray}
\begin{aligned}
\mbox{Sample-A:}
\end{aligned}
&&
\begin{aligned}
\hspace*{-4mm}
& u=0.57~\text{TeV}, &&
\hspace*{-2mm}
M_{\T}^{} \!\simeq M_{\T'}^{}\!=720\,\text{GeV},
&& \yt_S^{}=1.5, && \lambda_4^{}=0.4, && \lambda_5^{}=0.1, && \lambda_6^{}=0.08;
\end{aligned}
\label{eq:sample-A}
\qquad
\\
\begin{aligned}
  \mbox{Sample-B:}
\end{aligned}
&&
\begin{aligned}
\hspace*{-4mm}
& u=1.2\,\text{TeV}, &&
M_{\T}^{} \!\simeq M_{\T'}\!=720\,\text{GeV},
&& \yt_S^{} =0.88, &&
\hspace*{-1.7mm}
\lambda_4^{}=0.1, && \lambda_5^{}=0.3, && \lambda_6^{}=0.2;
\end{aligned}
\label{eq:sample-B}
\end{eqnarray}
\label{eq:samples}
\end{subequations}
\hspace*{-2mm}
where the couplings \,$(\yt_S^{},\, \lambda_4^{},\,\lambda_5^{},\,\lambda_6^{})$\,
are defined at a renormalization scale $\,\mu=M_S^{}$.\,
The DM mass $\,M_\X^{}$\, is irrelevant to
the analysis of vacuum stability and perturbativity.
But for the following LHC analysis and for the DM detection analysis in Sec.\,\ref{sec:5},
we will define $\,M_\X < M_S^{}/2\,$ in Sample-A and $\,M_\X > M_S^{}/2\,$ in Sample-B.
For the masses of heavy vector-like quarks, we choose a benchmark value
above the current lower bound on $M_{\T'}^{}$ (shown in Sec.\,\ref{sec:2}).
Our goal is to accommodate the observed excess of diphoton rate,
with reasonable $S$ Yukawa coupling which is consistent with the
requirements of stability and perturbativity.
Since $M_{\T}^{}$ and $M_{\T'}^{}$ include contribution from the bare mass
term ($M_0^{}$), the diphoton rate is not connected to the simple
ratio of mass $M_{\T}^{}$ and VEV $u$\,.\, Thus, we can properly choose $\,u$\,
according to the desired value of $\,\lambda_2^{}$\,.\,
We construct Sample-A and -B for different purposes here.
Sample-A has $\,M_\X < 375\,$GeV, which can produce both
the diphoton excess and the invisible decay $\,S\!\to\chi\chi$\,,\, and have
the vacuum instability bound much higher than TeV scale at the same time.
If assume no invisible decay, we derive a sizable diphoton cross section
at the LHC Run-2 as follows,
\beqa
\label{eq:CX-gaga-A}
\sigma_0^{}(pp \!\to\! S \!\to\! \gaga ) = 7.3\,\text{fb},
~~~~~ \text{(Sample-A)},
\eeqa
where the parton distribution function MSTW08 \cite{MSTW08} is used.
This is higher than the central value of fitted
diphoton signals\,\cite{fit} of Run-2 data\,\cite{ATLAS}\cite{CMS}.
After the invisible decay channel $\,S\!\to\X\X\,$ is open, we find
that for $\,M_\X^{}=100-350\,$GeV,\,
the diphoton cross section varies within the range,
\beqa
\label{eq:CX-gaga-A2}
\sigma(pp \!\to\! S \!\to\! \gaga ) \,=\, 3.1\!-4.9\,\text{fb},
~~~~~ \text{(Sample-A).}
\eeqa
This is consistent with the recent fit of combined
LHC Run-2 and Run-1 diphoton rate\,\cite{fit},
$\,\sigma[gg\to\XXX\to\gaga ]=4.6\pm 1.2\,$fb,\,
well within the $2\sigma$ range.\,
We will further analyze the invisible decay channel in Sec.\,\ref{sec:5.2}.
For Sample-B, we consider $\,M_\X > 375\,$GeV,\,
and optimize the parameters to accommodate both the diphoton excess
and the Higgs inflation, which maintains the vacuum stability and perturbativity
up to the inflation scale. Thus, we compute
\beqa
\label{eq:CX-gaga-B}
\sigma(pp \!\to\! S \!\to\! \gaga ) \,=\, 2.5\,\text{fb},
~~~~~ \text{(Sample-B),}
\eeqa
which is consistent with the recent fit of LHC diphoton rate\,\cite{fit}
and well within the $2\sigma$ range.\,
The non-observation of the 750\,GeV resonance in the di-jet channel ($S\!\to\! gg$)
so far could give important constraint.
In our model, for $\,\alpha\lesssim 10^{-3}$,\, we find,
\beqa
\sigma(pp \!\to\! S \!\!\to\! gg) \,\simeq\, 21\sigma(pp \!\to\! S \!\!\to\! \gaga )\,.
\eeqa
This is far below the CMS constraint on the di-jet cross section ($<1.8$\,pb)
by using the LHC Run-1 data\,\cite{CMSdijet}.
Another loop-induced channel is the rare decay $\,S\!\to Z\gamma$\,.\,
In our model, its cross section is much smaller than that of the diphoton final state,
and is well below the current constraint \cite{Aad:2014fha}.
For $\,S \!\to\! hh,t\bar{t},b\bar{b}\,$ decay modes,
the signal rates are negligibly small in the $\,\alpha\!\sim\! 10^{-3}$ region.
In this $\,\alpha\,$ range, $\,S \!\to\!WW,ZZ$\, are mainly induced
by the $\T\,(\T')$ triangle loops, and we find the following relations,
\beqa
\label{eq:VV-750}
\sigma(pp \!\to\! S \!\!\to\!WW)\simeq 4.3\sigma(pp \!\to\! S \!\!\to\!\gamma\gamma)\,, ~~~~~
\sigma(pp \!\to\! S \!\!\to\!ZZ)\simeq 0.65\sigma(pp \!\to\! S \!\!\to\!\gamma\gamma)\,.
\eeqa
Note that for all the loop induced decay modes, $\,S \!\!\to \gamma \gamma, gg, WW, ZZ$,\,
the relative sizes are fixed by the gauge quantum numbers of the heavy vector-like quarks
$(\T'\!,\,\T)$.\,
The LHC Run-1 data imposed upper limits on $WW$ and $ZZ$ final states \cite{8TeVVVlimit}.
These limits may be converted into bounds on the cross sections at the LHC Run-2\,(13\,TeV),
implying that the upper bounds on $WW$ and $ZZ$ cross sections around the 750\,GeV region
are roughly 220\,fb and 56\,fb, respectively \cite{Low:2015qep}.
They are far above our prediction \eqref{eq:VV-750} inferred from the
diphoton signals.

\section{Vacuum Stability and Higgs Inflation}
\label{sec:inflation}
\label{sec:4}

A principal motivation of our model is to ensure the stability
of Higgs potential in the very early universe when the scale of energy density
is much higher than the weak scale. We recall that, in general, a scalar coupling tends to
stabilize the potential while a Yukawa coupling tends to destabilize it.
In our model, this means in particular that the new Yukawa coupling $\,\yt_S^{}$
should not be too large. We find that Sample-B does meet this criterion to maintain
vacuum stability up to inflation scale.
In Sec.\,\ref{sec:4.1}, we first study the vacuum stability
for both samples A and B. Then, in Sec.\,\ref{sec:4.2}, we take the advantage of Sample-B
to realize successful Higgs inflation.

\subsection{Renormalization Group Running and Vacuum Stability}
\label{sec:4.1}
\vspace*{2.5mm}

 The vacuum stability may be studied by directly computing the effective Higgs potential with
 loop corrections, or by resumming up loop corrections into running couplings of
 the tree-level Higgs potential via renormalization group (RG).
 We will use the RG approach for the current analysis.
 Thus, we can apply the tree-level stability condition to the running scalar couplings
 and derive stability bound on the allowed running energy scale.
 For our model, the vacuum stability is mainly dictated by the competition of running contributions
 between scalar loops (involving scalar self-couplings) and fermion loops (involving
 the top and $\TT$ Yukawa couplings), since the contributions from gauge couplings
 $(g_s^{},\,g,\,g')$ are minor. Inspecting the Higgs potential \eqref{eq:V},
 we have the tree-level stability conditions,
\begin{equation}
\begin{aligned}
  & \lambda_{1,2,4}^{} \geqq 0\,,
  \hspace{5mm}\lambda_3^{}\geqq -2\sqrt{\lambda_1^{}\lambda_2^{}}\,,
  \hspace{5mm}\lambda_6^{}\geqq -2\sqrt{\lambda_1^{}\lambda_4^{}}\,,
  \hspace{5mm}\lambda_5^{}\geqq -2\sqrt{\lambda_2^{}\lambda_4^{}}\,,
\\
  & 2\lambda_1^{}\lambda_5^{}-\lambda_3^{}\lambda_6^{}\geqq
    -\sqrt{\(4\lambda_1^{}\lambda_2^{}\!-\lambda_3^2\)
     \(4\lambda_1^{}\lambda_4^{}\!-\lambda_6^2\)}\,.
\end{aligned}
\end{equation}

To further realize Higgs inflation, we consider the joint effective theory
which combines our model (Table\,\ref{tab:1}) with the general relativity,
and includes the unique dimension-4 non-minimal coupling term
$\,\xi R H^\dag H\,$,\, where $\,R\,$ is the Ricci scalar curvature.
As before\,\cite{HX},
we will use the SM two-loop $\beta$ functions together with the one-loop
$\beta$ functions of the non-minimal coupling $\,\xi\,$ and other couplings
involving new scalars and new fermions, including the $s$ factor which arises from
the non-minimal coupling term\,\cite{HX,Lerner2011,Allison2013}.
The two-loop $\beta$ functions of SM with $s$ insertions and
the one-loop $\beta$ function for $\,\xi\,$ were given in \cite{HX,Allison2013}.
We present the contributions to the $\beta$ functions
by the new couplings in Appendix\,B.
The one-loop matching at top mass is done as described in \cite{Degrassi2012}.

\begin{figure}[t]
\centering
\includegraphics[height=5.5cm,width=0.48\textwidth]{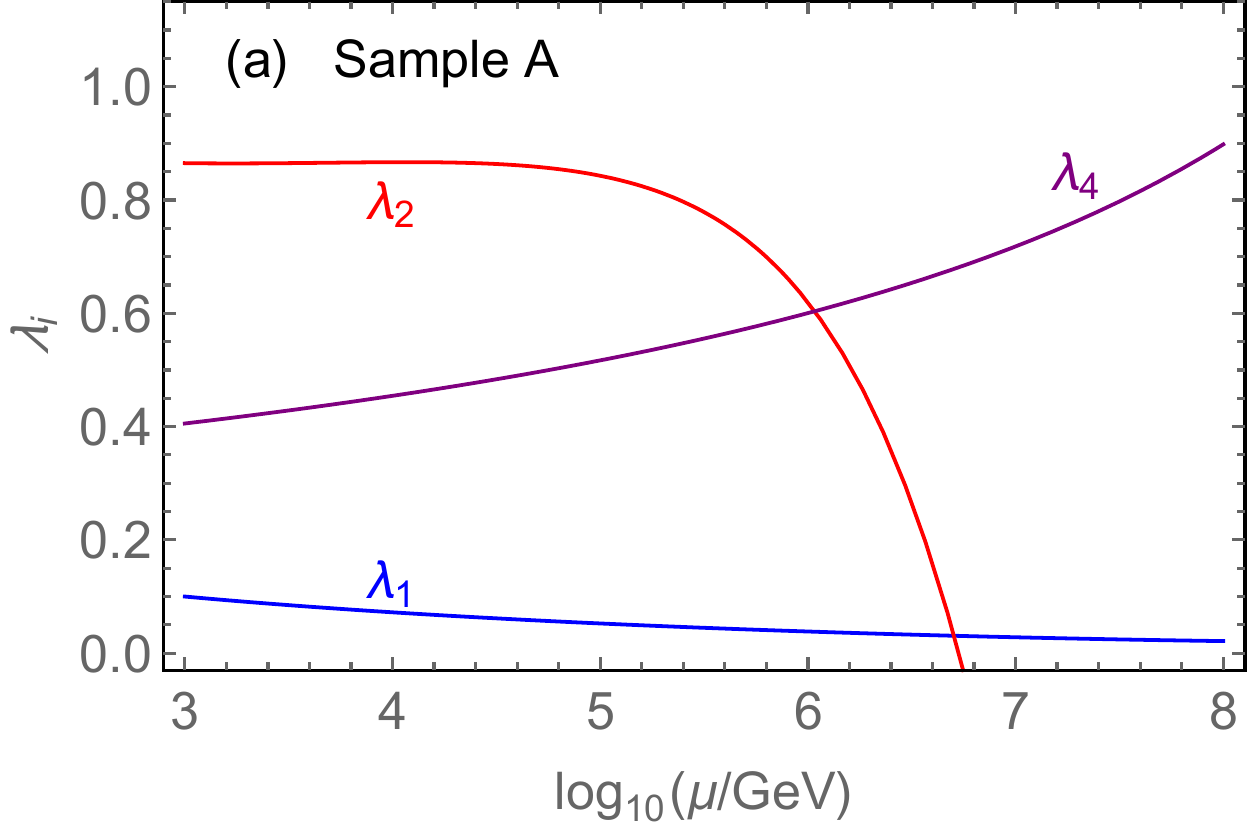}
\includegraphics[height=5.7cm,width=0.48\textwidth]{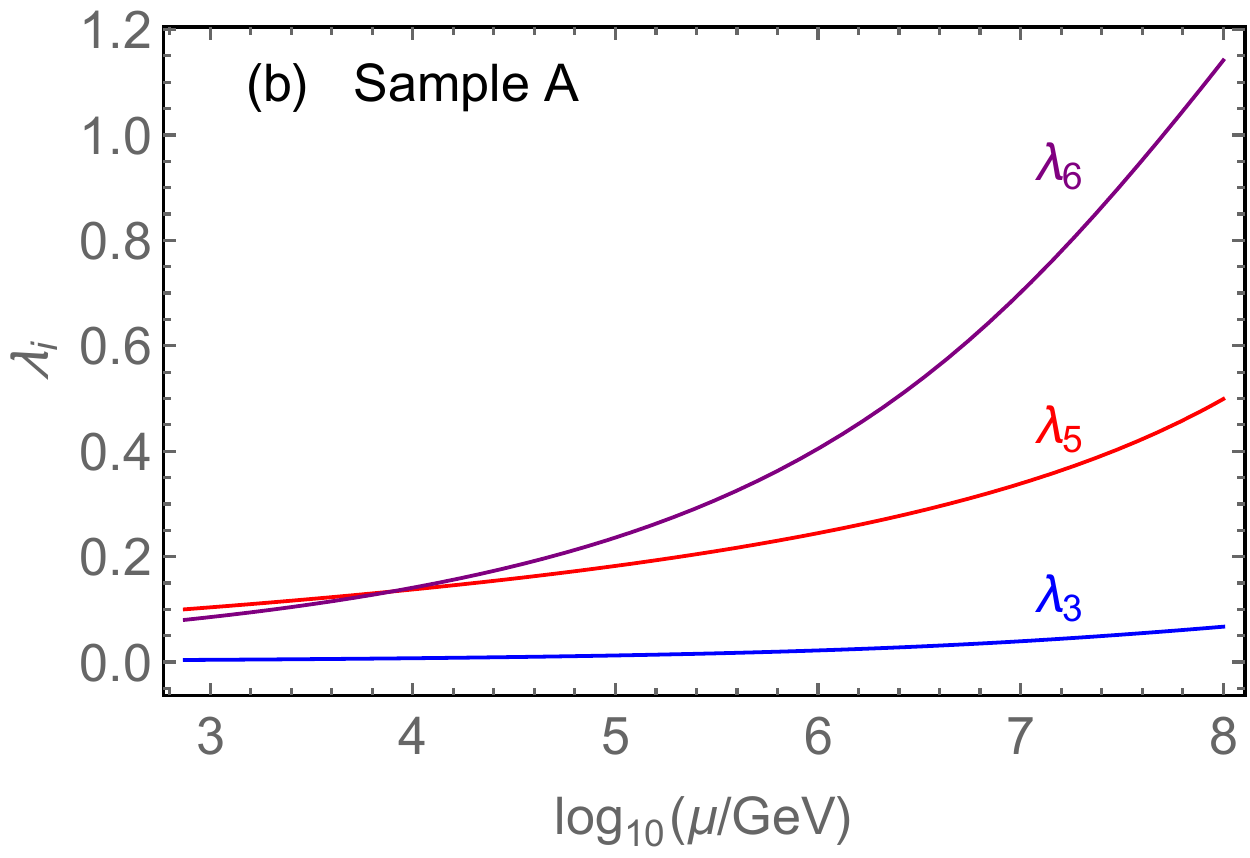}
\\
\includegraphics[height=5.7cm,width=0.48\textwidth]{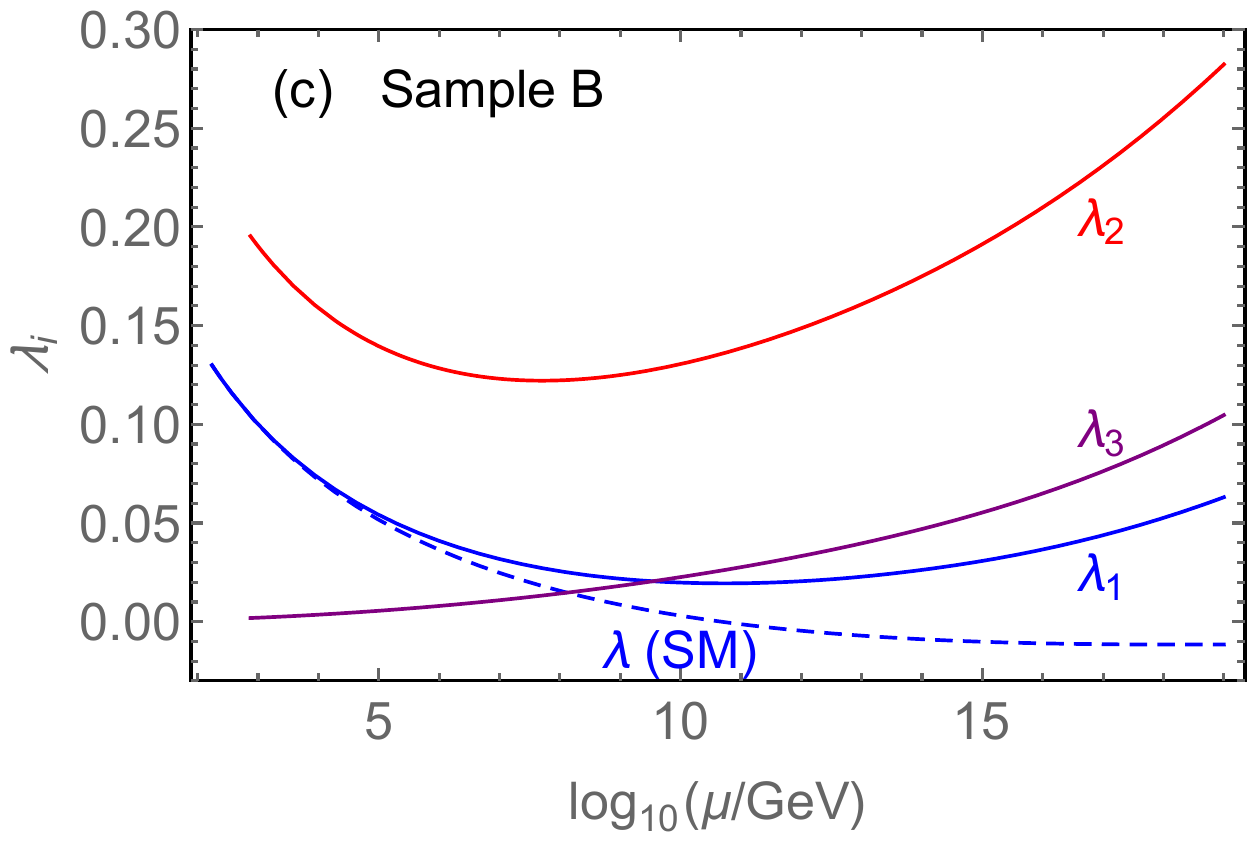}
\includegraphics[height=5.5cm,width=0.48\textwidth]{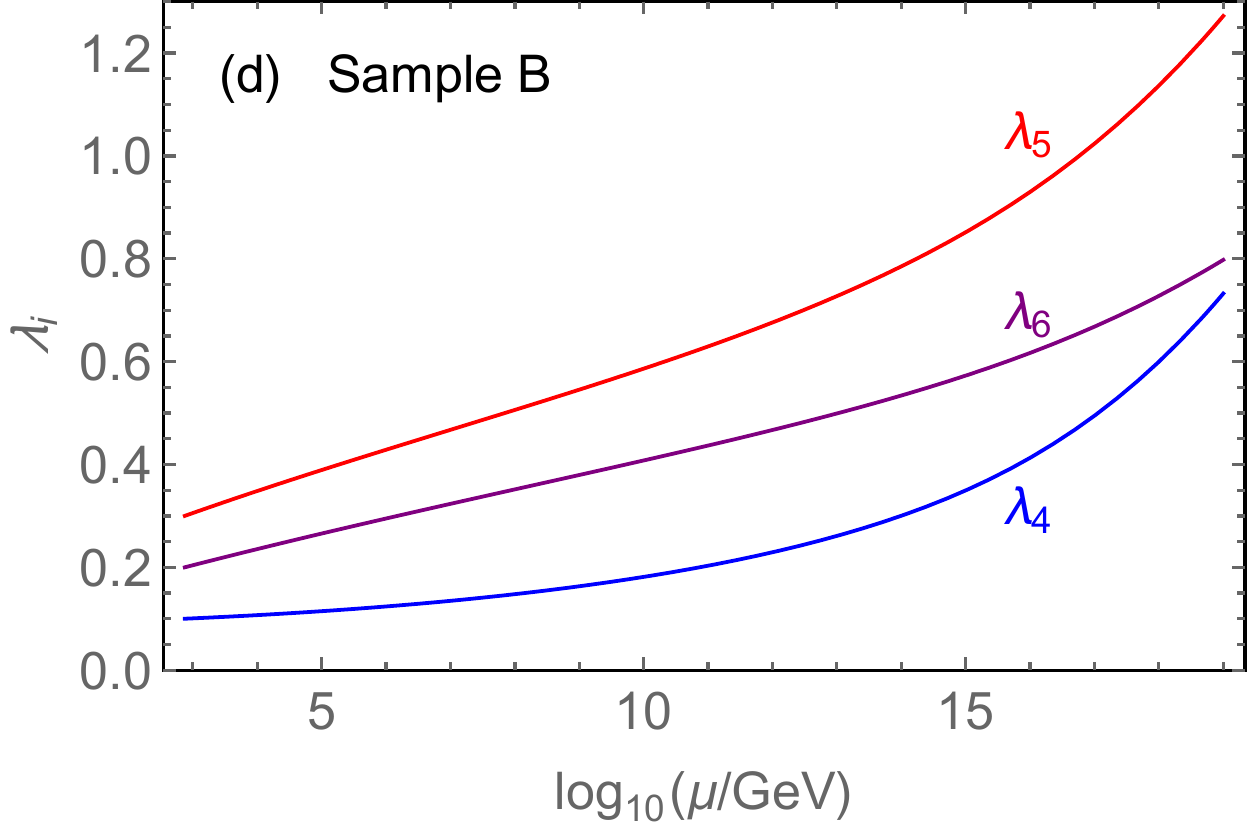}
\caption{Running scalar couplings $\,\lambda_j^{}\,$ of the present model
as functions of the RG scale $\,\mu$\,.\, Plots (a)-(b) show the running behaviors
for Sample-A, while plots (c)-(d) depict that for Sample-B.
The dashed curve in plot (c) represents running of the SM Higgs self-coupling
$\,\lambda\,$ up to two-loop RG.}
\label{lambda}
\label{fig:2}
\end{figure}

With these, we analyze the RG runnings for Sample-A and Sample-B, and
derive the vacuum stability bounds for the Higgs potential.
In Fig.\,\ref{fig:2}(a)-(b) and Fig.\,\ref{fig:2}(c)-(d), we present the running scalar couplings
as functions of the renormalization scale $\,\mu\,$.\,
For Sample-A, we find that the stability bound is reached around
$\,\mu\simeq 5.4\times\! 10^3$\,TeV, due to the decrease of $\,\lambda_2^{}$,\,
as shown in Fig.\,\ref{lambda}(a).
For Sample-B, we find that all scalar couplings remain positive
and perturbative up to Planck scale.
This ensures the Higgs potential to be a valid description
of the inflation potential, so the inflation trajectory is stable against
small perturbations in the directions of $\,(S,\,\chi)$\,.\,

In both Samples A and B, we fix the singlet mass as $\,M_S^{}=750$\,GeV.\,
It would also be instructive to inspect how the vacuum stability in our samples
is affected by varying the singlet mass around $\,M_S^{}=750$\,GeV.\,
We can clearly check this effect through the one-loop $\beta$ functions
(\ref{betafunc1}) and (\ref{betafunc2}). The main concern of the Higgs vacuum stability
in the SM model is the running of $\,\lam_1^{}$,\, which may drive $\,\lam_1^{}$\,
into negative value at high energies. This problem is relieved in our model
due to the positive contribution of $\,\beta_{\lam_1}^{}\,$ from
$\,\lam_3^{}\,$ and $\,\lam_6^{}$.\,
The main contribution actually comes from $\,\lam_6^{}\,$,\,
since the smallness of mixing angle $\al$ requires $\lam_3^{}$ be tiny,
as can be seen from Eq.\,(\ref{eq:lambda123}).
On the other hand, since $\,\lam_3^{}\,$ is tiny, from \eqref{eq:massH} we see that
varying $M_S^{}$ mainly affects the value of $\,\lam_2^{}$.\,
But, Eq.\,(\ref{betafunc2}) shows that the one-loop $\,\beta_{\lam_6}^{}\,$
does not depend on $\,\lam_2^{}\,$ explicitly. So it is clear that the vacuum stability
is very insensitive to the variation of $M_S^{}$.\,
We also check this numerically by varying $M_S^{}$ in Sample\,B,
and find that the vacuum stability is well preserved over the mass range
$\,700\,\text{GeV}\leqq M_S^{}\leqq 800$\,GeV (with other parameters fixed).

\subsection{Realizing Higgs Inflation}
\label{sec:4.2}
\vspace*{2.5mm}

In Higgs inflation, it is the Higgs field that successfully drives the cosmic inflation,
and the same Higgs field will spontaneously break the electroweak gauge symmetry at low energies.
The typical energy density scale during Higgs inflation is around $10^{16}\,$GeV.\,
Hence, for our model to hold consistently up to the inflation scale, the RG running will play
an essential role. We have done the RG running analysis in Sec.\,\ref{sec:4.1}.
Fig.\,\ref{fig:2}(c)-(d) shows that Sample-B is a possible candidate for realizing
successful Higgs inflation. In this subsection, we will apply this to directly derive
the Higgs inflation potential and inflationary observables.

\begin{figure}[t]
\centering
\includegraphics[width=0.60\textwidth]{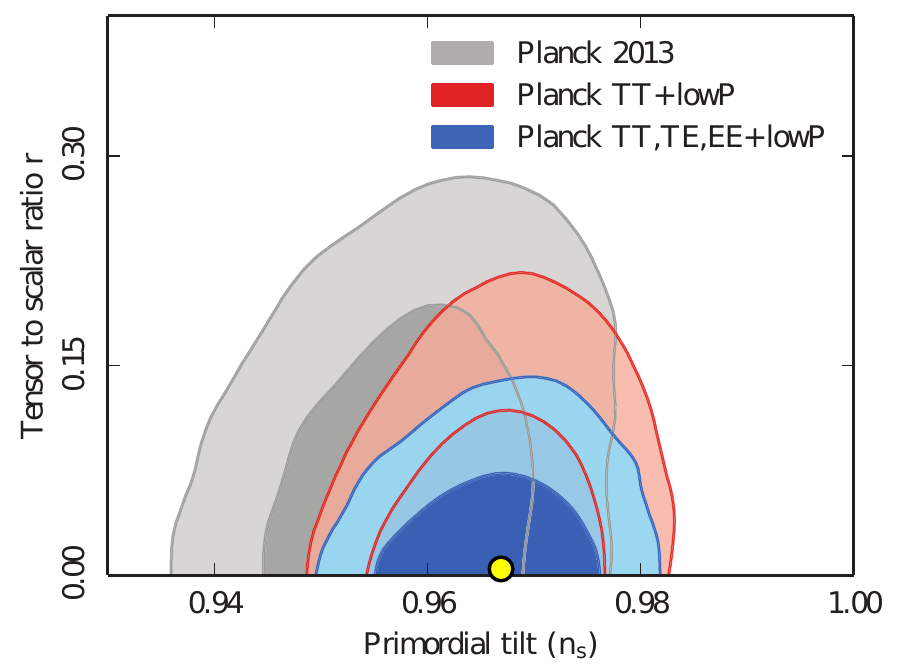}
\vspace*{-3mm}
\caption{Predicted scalar tilt $\,n_s^{}\,$ and tensor-to-scalar ratio $\,r\,$
by the present model (shown as the yellow dot), in comparison with various
Planck limits in 2015 and 2013 \cite{Planck2015Inf}. In each type of contours,
the shaded darker and lighter regions represent the 68\% and 95\% limits, respectively.
}
\label{rvsns}
\label{fig:3}
\end{figure}

In Higgs inflation, the unique non-minimal coupling term
$\,\xi R H^\dag H\,$ plays the key role to flatten the Higgs potential
at high energies. It is conventional to work in the Einstein frame, and
we find that the Higgs inflation occurs along the valley of the Higgs potential
where the fields $(S,\,\X)=(0,\,0)$.\,
In Einstein frame, we express the Higgs field $\,h\,$ in terms of
canonically normalized field $\,\varphi\,$,\, through
$\,\di\varphi/\di h=(\Omega^2\!+\!6\xi^2h^2/\Mp^2)^{1/2}/\Omega^2$.\,
Here $\,\Omega^2\!=1\!+\xi h^2/\Mp^2$\,
is the Weyl factor that brings the action from its defining (Jordan) frame
to the Einstein frame, and $\,M_{\text{Pl}}^{}\simeq 2.4\times\! 10^{18}$\,GeV
is the reduced Planck mass. Thus, we can rewrite the Higgs potential in terms of
the normalized field $\,\varphi\,$,\, and expand it in the large field region
($h\gg \Mp/\xi$\,),
\beqa
V(\varphi) \,\simeq\, \frac{\,\lambda_1^{} M_{\text{Pl}}^2\,}{4\,\xi^2}
\bigg(1-e^{-\sqrt{2/3}\,\varphi/M_{\text{Pl}}}\bigg)^2,
\eeqa
From this potential, we can directly compute the first two slow-roll parameters,
$\,\epsilon\,$ and $\,\eta\,$, as well as the number of $e$-foldings $\,N_e\,$
between the beginning and end of the observable inflation,
\beqa
\epsilon \,=\,\frac{\,M_{\text{Pl}}^2\,}{2}\frac{\,V_{\varphi}^{\prime 2}\,}{V^2} \,,
\hspace*{8mm}
\eta \,=\, M_{\text{Pl}}^2\frac{\,V''_{\varphi}\,}{V} \,,
\hspace*{8mm}
N_e \,=\, \FR{1}{\Mp}\!\!\int_{\varphi_{\text{end}}^{}}^{\varphi_{0}^{}}
\FR{\di\varphi}{\sqrt{2\ep\,}\,}\,.
\eeqa
Here we use $\,\varphi_{0}^{}$ and $\,\varphi_{\text{end}}^{}\,$
to denote the values of the inflaton field $\,\varphi\,$
at the beginning and end of the observable inflation.
The condition for ending the inflation is given by $\,\ep<1\,$ in our model,
and the beginning of observable inflation can then be determined by the
needed number of $e$-foldings, which is $\,N_e\simeq 59\,$
for typical Higgs inflation \cite{Reheating}.
Then, we derive the inflation observables, including
the scalar amplitude $\,(V/\epsilon)^{1/4}$,\,
the scalar tilt $\,n_s^{}=1-6\eta+2\epsilon$\,,\,
and the tensor-to-scalar ratio $\,r=16\epsilon\,$,\,
at $\,\varphi=\varphi_{0}^{}$\,.\,
The inflation potential $\,V(\varphi)$\, contains one free parameter,
the non-minimal coupling $\,\xi$\,,\,
and it can be fixed by the Planck normalization for the scalar amplitude
$\,(V/\epsilon)^{1/4}$ \cite{Planck2015Amp}.
In our Sample-B, this corresponds to $\,\xi\simeq 8000$\,.\,
Then, we derive the predicted scalar tilt $\,n_s^{}\,$ and tensor-to-scalar ratio $\,r\,$,
\beqa
n_s^{}\,\simeq\, 0.967\,, ~~~~~~
r \,\simeq\, 0.004 \,, ~~~~~~~~(\text{Sample-B}).
\eeqa
We compare these predictions with the announced Planck limits \cite{Planck2015Inf}
and find good agreement. This comparison is presented in Fig.\,\ref{fig:3}.

\section{Realizing Dark Matter: Relic Abundance and Searches}
\label{sec:DM}
\label{sec:5}

In this section, we analyze the realization of the CP-odd singlet
$\,\X\,$ as the DM candidate. We furher study
its searches at the LHC and its (in)direct detections.
Sec.\,\ref{sec:5.1} analyzes the relic abundance for the DM $\,\X\,$,\,
and derives nontrivial constraints on the DM mass for both Sample-A and Sample-B.
Then, Sec.\,\ref{sec:5.2} studies the DM collider signals via invisible decays
of the 750\,GeV new resonance \,($S\!\to\!\X\X$)\, for Sample-A at the LHC Run-2.
Finally, we analyze the DM direct and indirect detections
in Sec.\,\ref{sec:5.3}--\ref{sec:5.4}.

\vspace*{1mm}
\subsection{Dark Matter Relic Abundance}
\label{sec:5.1}
\vspace*{2.5mm}

The singlet pseudoscalar $\,\X$\, serves as the DM candidate in our model.
It only couples to scalar particles via the Higgs potential \eqref{eq:V}.
For the DM annihilation processes, if the intermediate particle is $S$,\,
the final state particles can be $\,gg$, $hh$,
$hS$,\, $SS$,\, $\T\T$\, and $\T'\T'$,\, depending on whether the DM mass
is large enough to open the relevant channels. On the other hand, if the
intermediate particle is the light Higgs boson $\,h\,$,\,
then $\,\chi\chi\,$ will annihilate into SM particles and $SS$.
For our analysis, we summarize the nonzero coupling constants of relevant vertices
around $\,\alpha\sim 0$\, region,
%
\begin{eqnarray}
  \lambda_{\chi \chi S}^{}
=
- \lambda_5^{} u \,,
\,\,\,
  \lambda_{\chi \chi h}^{}
=
- \lambda_6^{} v \,,
\,\,\,
  \lambda_{\chi \chi hh}^{}
=
- \lambda_6^{} \,,
\,\,\,
  \lambda_{\chi\chi SS}^{}
=
- \lambda_5^{} \,,
\,\,\,
  \lambda_{hhh}
=
- 6 \lambda_1^{} v \,,
\,\,\,
  \lambda_{hhS}^{}
=
- \lambda_3^{} u \,,
\,\,\,
  \lambda_{hSS}^{} = - \lambda_3^{} v \, .
\end{eqnarray}
\label{eq:scalar-couplings}
%
\hspace*{-2mm}
The only parameter unspecified in our samples (\ref{eq:samples}) is $\,\mu_3^{}\,$,\,
which is connected to the DM mass $M_\X^{}$\,.\,
Since it is irrelevant to the stability analysis in Sec.\,\ref{sec:4},
we treat it as a free parameter corresponding to the DM mass.

\begin{figure}[t]
\vspace*{-3mm}
\centering
\includegraphics[height=6cm,width=7.7cm]{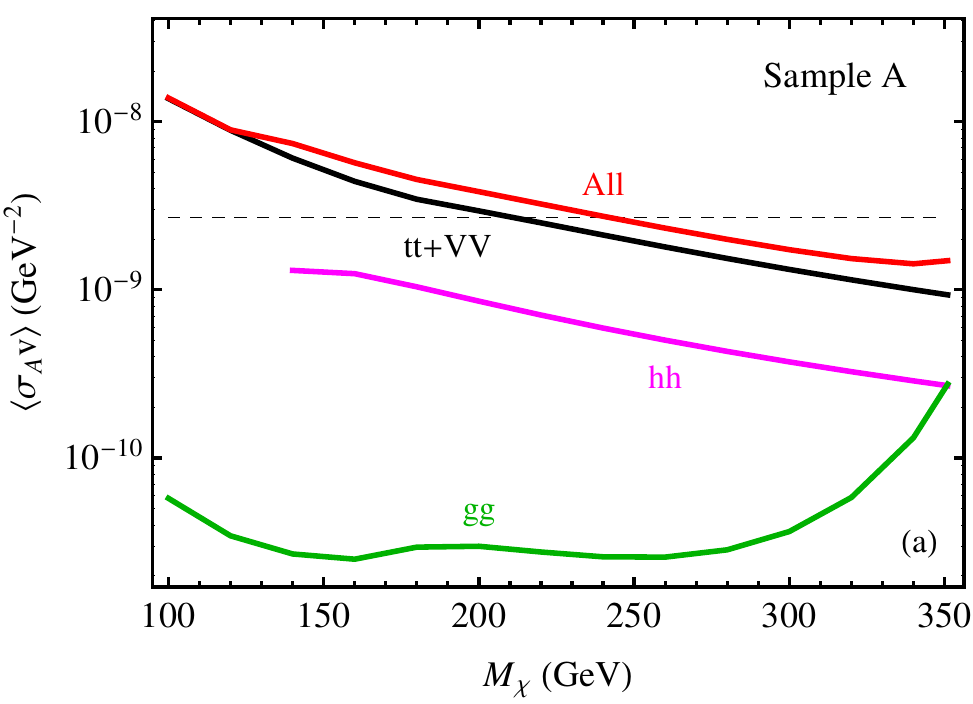}
\quad
\includegraphics[height=6cm,width=7.7cm]{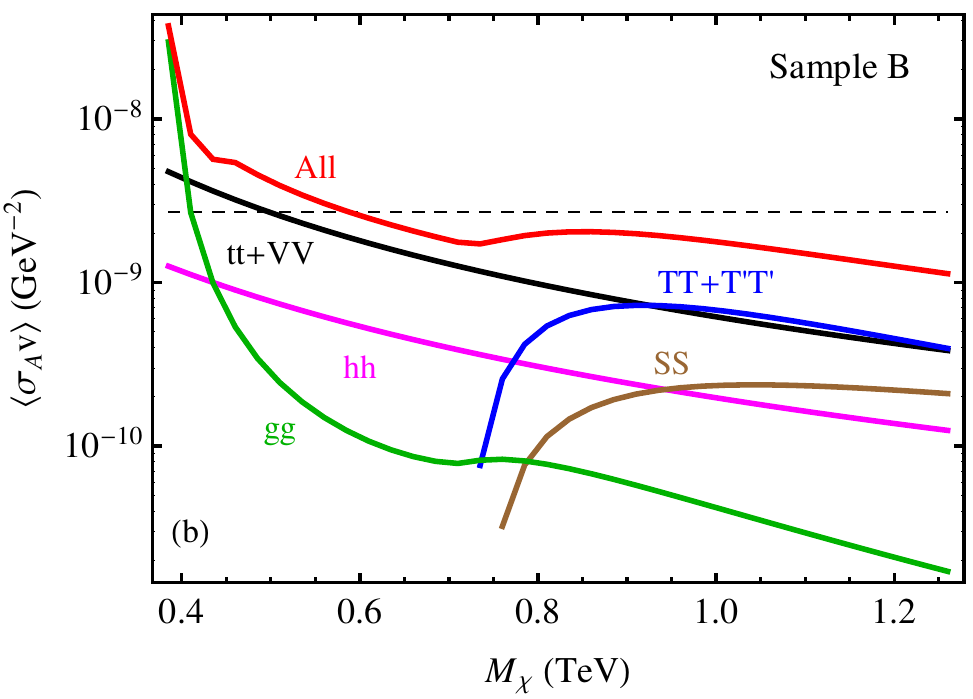}
\vspace*{-2.5mm}
\caption{Thermal averaged cross section for DM annihilation, as a function of the DM mass
$\,M_\X^{}\,$ for Sample-A [plot-(a)] and Sample-B [plot-(b)].}
\label{fig:RDAB}
\label{fig:4}
\end{figure}

We compute the thermal averaged cross sections for DM annihilations.
In Figs.\,\ref{fig:4}(a)-(b), we present them as functions of the DM mass $\,M_\X^{}\,$
for Sample-A and Sample-B. The black dashed line shows the typical cross section
$\,\langle\sigma_Av\rangle = 2.7 \times\! 10^{-9}\,\mbox{GeV}^{-2}$,\,
which corresponds to the observed DM relic density.
The solid curves represent our theory prediction.
For Sample-A, we consider the lighter mass region $\,\MX < 375\,$GeV,\, as shown in
\gfig{fig:4}(a). With the sample input $\,\lambda_{\X\X S}^{}\!=-57\,$GeV\.,\,
the annihilation cross section is dominanted by $\,t\bar{t}\,$ and $\,VV\,$ final states.
They come from the Higgs exchange and are sensitive to $\,\lambda_6^{}\,$.\,
To produce the observed relic density, we find that the typical DM mass is around
$\,\MX =240$\,GeV\,.\, For Sample-B, we consider the larger mass range
$\,\MX > 375\,$GeV\,,\, as shown in
\gfig{fig:4}(b). With the sample input $\,\lambda_{\chi\chi S}=-360\,$GeV,\,
$t\bar{t}$ and $VV$ channels still give the main contribution.
The annihilation cross sections of $\,\X\X\to\T\T,\T'\T'\,$ become barely comparable
when $\,\MX >M_{\T(\T')}^{}$.\, To generate the observed DM relic density,
we find the DM mass around $\,\MX\! =588$\,GeV,\,
which is mainly determined by $\,t\bar{t}\,$ and $\,VV\,$  cross sections.

\vspace*{1mm}
\subsection{Dark Matter Production at the LHC Run-2}
\label{sec:5.2}
\vspace*{2.5mm}

\begin{figure}[t]
\centering
\includegraphics[height=6cm,width=7.99cm]{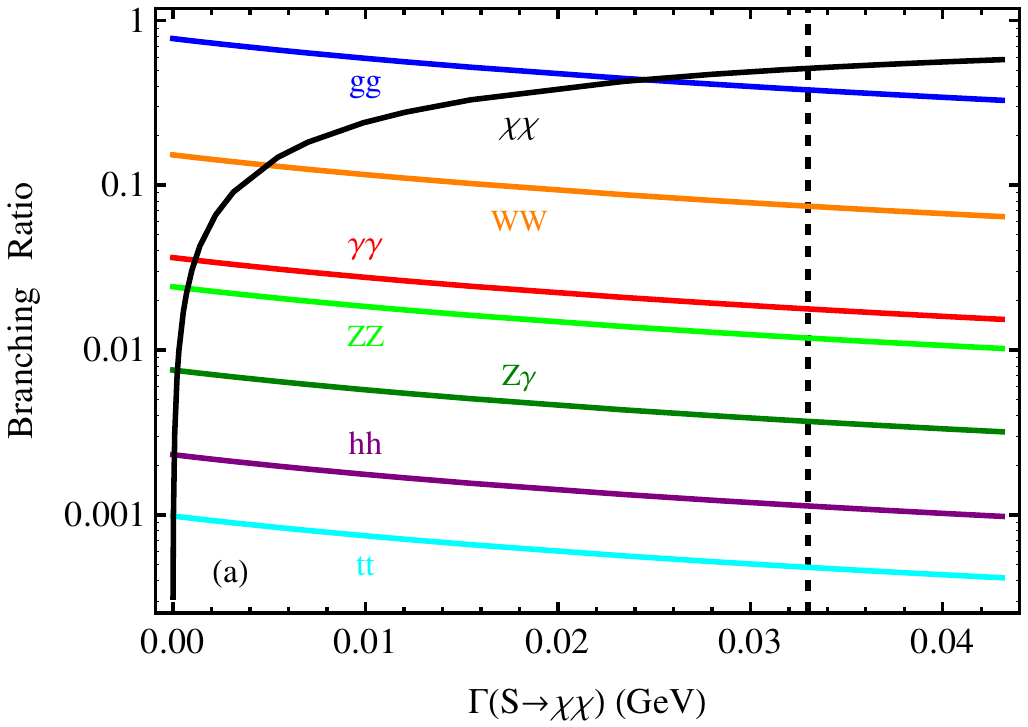}\quad
\includegraphics[height=6cm,width=7.6cm]{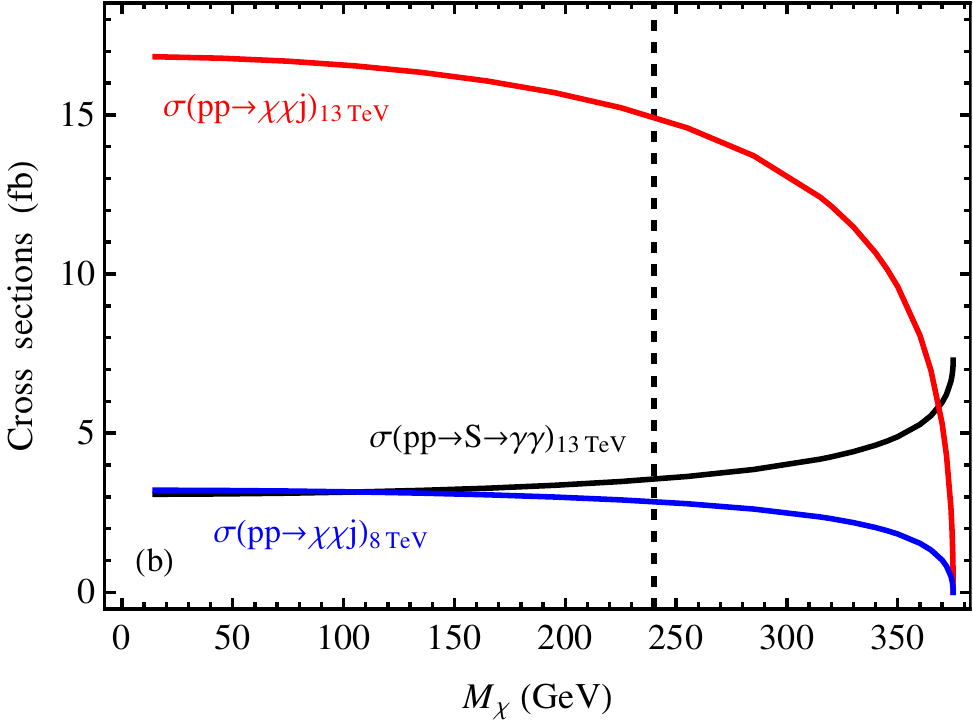}
\vspace*{-2mm}
\caption{(a).~Branching ratios of $\,S\,$ decay as functions of invisible width
$\,\Gamma (S\!\to\!\X\X)\,$.\,
(b).~DM production cross section with mono-jet as a function of mass $\,\MX\,$.\,
The red and blue curves present the parton level mono-jet cross sections at the LHC\,(13TeV) and
the LHC\,(8TeV), respectively, under preselection cuts: $p_{Tj}^{} > 120\,$GeV,\,
$|\eta_j^{}| < 2\,$,\, and $\,E^{\text{miss}}_T > 150\,$GeV.\,
The black curve shows the diphoton cross section $\,\sigma(pp\!\to\! S\to\!\gamma\gamma)\,$
at the LHC\,(13TeV). The vertical dashed line denotes the demanded DM mass for realizing
the observed thermal relic density.}
\label{fig:BRdm}
\label{fig:5}
\end{figure}

An important case is that the DM mass falls into
the region $\,\MX < M_S^{}/2$\,,\, as described by Sample-A,
so the $750$\,GeV new resonance has significant invisible decays
$\,S \to \chi\chi$\,.\,
The invisible decay width of $\,S$\, is
\begin{eqnarray}
\Gamma(S\!\!\to\!\chi\chi)
\,=\, \frac{\lambda_{\chi\chi S}^2}{\,32\pi M_S^{}\,}\!
\sqrt{1\!-\!\frac{\,4M^2_\X\,}{M_S^2}\,}\, .
\end{eqnarray}
In \gfig{fig:5}(a), we present the branching fractions of $S$ decays as functions
of the invisible width $\,\Gamma (S\!\to\!\X\X)\,$.\,
To generate the right amount of DM relic aboundance requires
$\,\MX =240\,$GeV,\, as shown in Fig.\,\ref{fig:4}(a). This is marked by the
black vertical dashed line in \gfig{fig:5}(a). At $\,\MX\! =240\,$GeV\,,\,
the corresponding invisible width and branching fraction are,
$\,\Gamma (S\!\!\to\!\X\X)=0.033\,$GeV and
$\,\text{Br}(S\!\!\to\!\X\X)=51\%$.\,
Under narrow width assumption, the diphoton cross section is related to
$\,\text{Br}(S\!\!\to\!\X\X )\,$ as follows,
\begin{eqnarray}
\sigma(pp\!\to\! S\!\!\to\!\gamma\gamma) \,=\,
\sigma_0^{}(pp\!\to\! S\!\!\to\!\gamma\gamma)
\left[1-\textrm{Br}(S\!\!\to\!\X\X )\right] ,
\end{eqnarray}
where $\,\sigma_0^{}(pp\!\to\! S\!\!\to\!\gamma\gamma)=7.3\,$fb\,
is the cross section at 13\,TeV assuming zero invisible decay width,
as given in Eq.\,\eqref{eq:CX-gaga-A} for Sample-A.
The diphoton cross section is depicted by the black solid curve as a function of
$\,\MX\,$ in \gfig{fig:BRdm}(b). The vertical dashed line denotes
$\,\MX\! =240\,$GeV,\,
at which we have
\beqa
\sigma(pp\!\to\! S\!\!\to\!\gamma\gamma)\simeq 3.6\,\text{fb}\,,
\hspace*{10mm} (\text{Sample-A})\,.
\eeqa
In passing, Ref.\,\cite{fermionDM} studied invisible decays of the 750GeV resonance into
a pair of Dirac fermion DM in a simplified DM model .

The LHC can probe the invisible decay $\,S \to \chi\chi$\, of our model
via mono-jet searches. When $S$ is produced by $gg$ fusion,
an extra gluon can be radiated from either the initial gluons or the heavy quarks in the loop.
This channel has been studied by ATLAS at 8\,TeV \cite{monojet}.
We generate $\,pp \!\to\! j\,\chi\chi\,$ events with $\,\MX = 240\,$GeV\,
by using {MadGraph5}$\_\,${aMC@NLO} \cite{MG5aMC}\cite{Hirschi:2015iia},
and apply the preselection cuts, $\,p_{Tj}^{} > 120\,$GeV,\,
$\,|\eta_j^{}| < 2\,$,\, and $\,E^{\text{miss}}_T > 150\,$GeV\,.\,
The cross section is \,2.8\,fb at the LHC\,(8TeV),\,
which only produces $57$ events with $20\,\mbox{fb}^{-1}$ integrated luminosity.
On the other hand, the uncertainty is still quite large. Even for the most sensitive
signal region SR9 with $E^{\text{miss}}_T > 700\,\mbox{GeV}$, our signal is about
the same size as the uncertainty.
Since only preselection cuts are applied in our simulation,
the signals are small enough to evade the current bound.
The parton level cross section under the preselection cuts for
$\,\sqrt s = 13\,\mbox{TeV}$\, and \,$\sqrt s = 8\,\mbox{TeV}$\,
are depicted by the red and blue curves in Fig.\,\ref{fig:5}(b).
We see that compared to the case of $\,\sqrt{s} =8$\,TeV,\,
the mono-jet cross section at $\,\sqrt{s} = 13\,\mbox{TeV}$\,
is about 5 times larger,
$\,\sigma(pp\to j\,\chi\chi)\simeq 15\,$fb\, for $\,\MX =240\,$GeV.\,
This signal may become observable with higher integrated luminosity at the Run-2.
For Sample-B with $\,\MX >375\,$GeV,\, a $\,\chi\chi\,$ pair could only be produced
via the off-shell exchange of $\,S$.\,
This makes the mono-jet cross section even smaller.

\vspace*{1.5mm}
\subsection{Dark Matter Direct Detection}
\label{sec:5.3}
\vspace*{2.5mm}

\begin{figure}[t]
\vspace*{3mm}
\centering
\includegraphics[height=6cm,width=7.7cm]{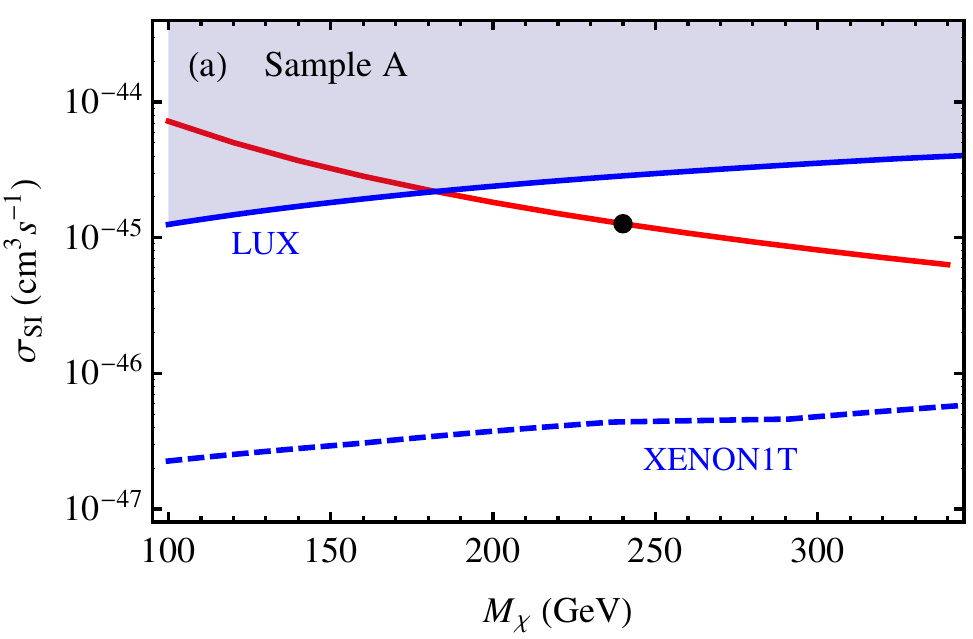}\quad
\includegraphics[height=6cm,width=7.7cm]{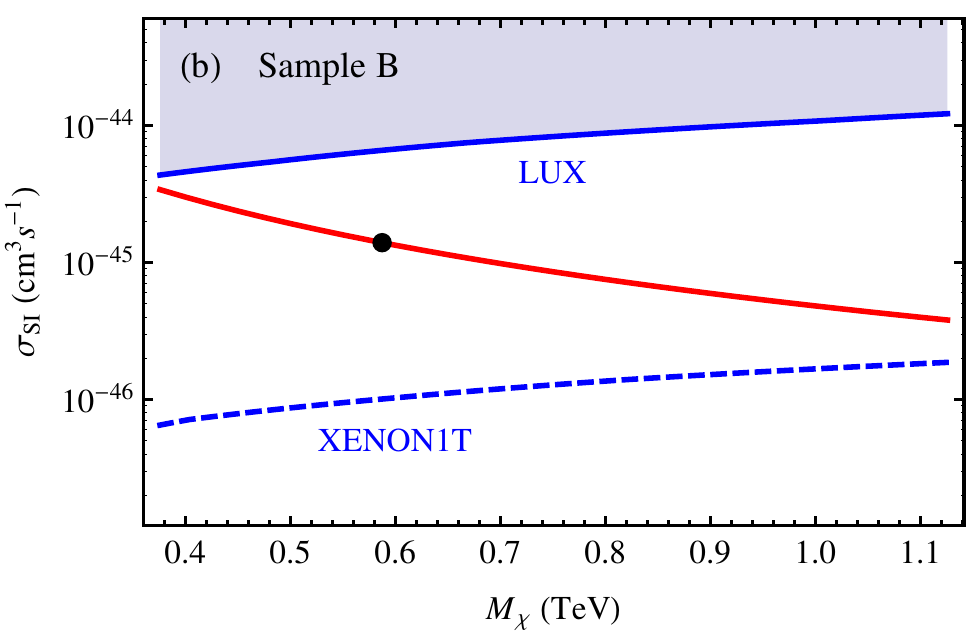}
\vspace*{-2mm}
\caption{DM-nucleon spin-independent cross section as a function of
$\,\MX\,$ for Sample-A [plot-(a)] and Sample-B [plot-(b)]. In each plot,
the red curve presents our prediction, and the black dot is dictated
by further imposing the constraint of the observed DM relic density.
The shaded region is excluded by the LUX measurements,
and the region above the blue dashed curve will be probed
by the upcoming Xenon1T experiment.}
\label{fig:DDDM}
\label{fig:6}
\vspace*{3mm}
\end{figure}

In the present model, DM interacts with the nucleon via exchanging both the Higgs boson
$h\,(125\text{GeV})$ and the new particle $S(750\,\text{GeV})$.\,
In the $\,\alpha\sim 0\,$ region, $\,h$\, interacts with the quark and gluon content
in the nucleon, while $\,S\,$ only interacts with the gluons via heavy quark triangle-loops.
The DM-nucleon interaction is spin-independent. We derive the DM recoil cross section,
\begin{eqnarray}
\sigma_{\textrm{SI}}^{} &\!\!=\!\!&
\frac{m_N^2}{\pi(M_\chi+m_N)^2}\left(G_{h,N}+G_{S,N}\right)^2,
\nonumber\\
G_{h,N}^{} &\!\!=\!\!&
\frac{\,\lambda_{\chi\chi h}^{}f_N^{}m_N^{}\,}{2v M_h^2},
\quad~~
G_{S,N}^{}=
\frac{2\lambda_{\chi\chi S}^{} \yt_S^{} m_N^{}\,}
{27 \sqrt{2}M_\mathcal{T} M_S^2}\left(1-\sum_{q=u,d,s}\!\! f_{N,q}\right) ,
\end{eqnarray}
where the nucleon mass $\,m_N^{}\!=0.939\,$GeV\, is the averaged mass of proton and neutron.
For the effective form factor, we use $\,f_N^{}=0.345$ \cite{Cline:2013gha},
and $\,(f_{N,u}^{},\,f_{N,d}^{},\,f_{N,s}^{})=(0.014,\,0.036,\,0.118)$\, \cite{Ellis:2000ds}.
In \gfig{fig:6}, we present the spin-independent cross section as a function of the DM mass
$\,\MX\,$ for Sample-A and Sample-B by the red curves.
Note that the contribution from $S$-exchange is heavily suppressed by
$\,M_S^{}=750\,$GeV.\, Given the cubic couplings
$(\lambda_{\chi\chi h}^{},\, \lambda_{\chi\chi S}^{})$
in Sample-A and Sample-B, we find that the DM-nucleon cross section is dominated
by $h$-exchange. The black dot denotes our prediction by imposing the constraint
of observed thermal relic density. Currently, the strongest constraint on the spin-independent
cross section comes from LUX experiment\,\cite{LUX2013} with the shaded region excluded
at 90\%\,C.L. Figs.\,\ref{fig:6}(a)-(b) show that our prediction (black dot) is currently viable.
But it is within the reach of the projected sensitivity of the upcoming Xenon1T\,\cite{XENON1T},
as represented by the blue dashed curve.

\vspace*{1.5mm}
\subsection{Dark Matter Indirect Detection}
\label{sec:5.4}
\vspace*{2.5mm}

\begin{figure}[t]
\centering
\includegraphics[height=6cm,width=7.7cm]{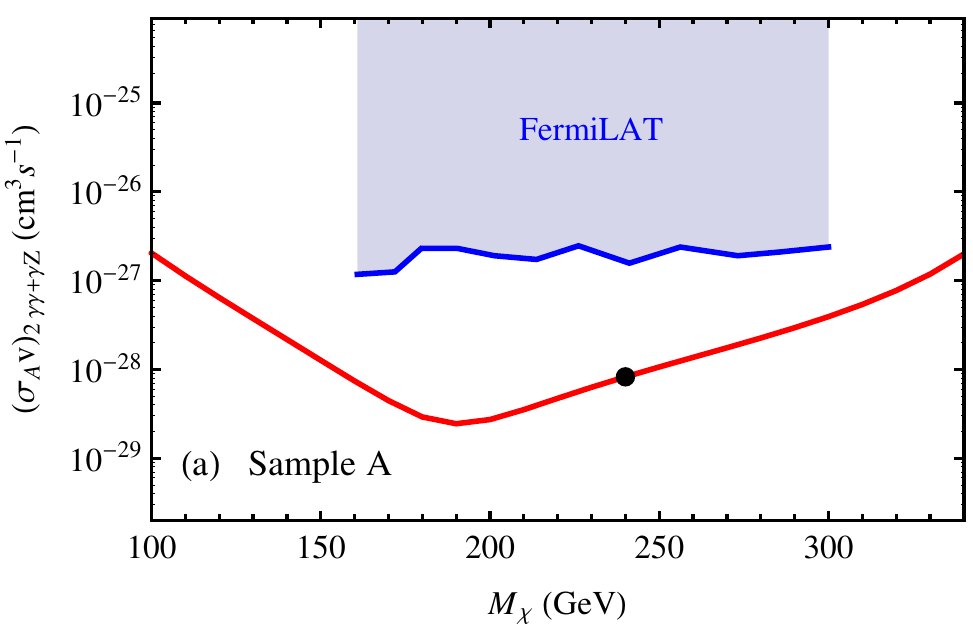}\quad
\includegraphics[height=6cm,width=7.9cm]{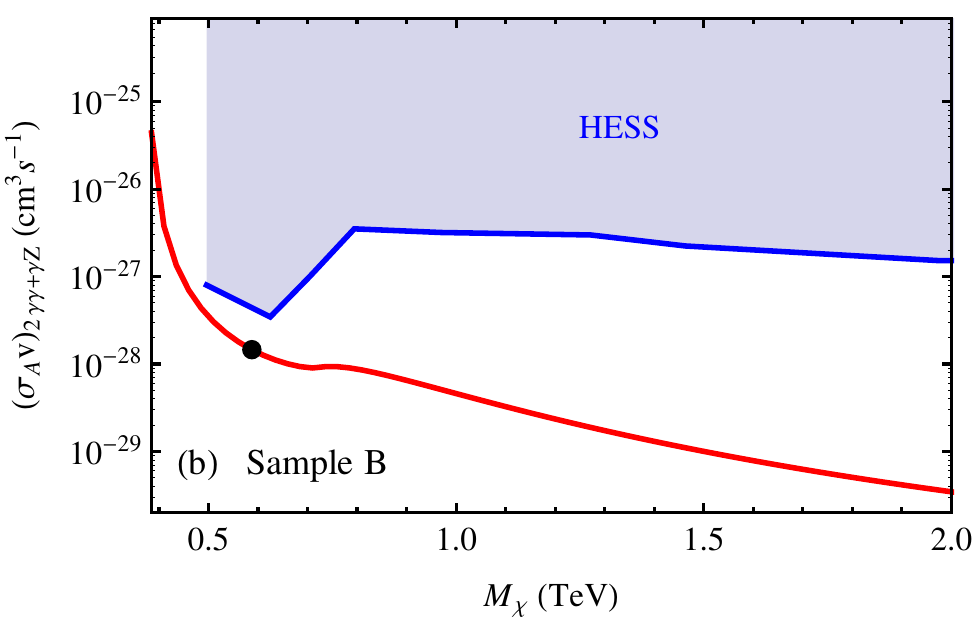}
\vspace*{-2mm}
\caption{Prediction of $\,2(\sigma_A^{}v)_{\gamma\gamma}^{}\!+(\sigma_A^{}v)_{\gamma Z}^{}$\,
as a function of $\,\MX\,$ for Sample-A [plot-(a)] and Sample-B [plot-(b)].
In each plot, the red curve presents the theory prediction, and the black dot is
dictated by further imposing the constraint of the observed DM relic density.
The shaded regions in plots (a) and (b) are excluded by Fermi-LAT and HESS experiments,
respectively.}
\label{fig:IDDM}
\label{fig:7}
\end{figure}

The DM annihilations can be also probed via indirect detections in the sky.
The first type is the gamma-ray spectral lines that arise from the DM annihilation
$\,\chi\chi\to\gamma X\,$,\, where $X$ denotes any other possible SM bosons.
In the present model, we have annihilation processes
$\,\chi\chi\to\gamma\gamma$\, and $\,\chi\chi\to\gamma Z\,$.\,
In the parameter region of $\,\alpha\sim 0$\,,\, we derive
the annihilation cross sections,
\beqs
\begin{eqnarray}
(\sigma_A^{} v )_{\gamma\gamma}^{} \hspace{-2mm} &\!=\!& \hspace{-2mm}
\frac{\,\alpha^2M_\chi^2}{4\pi^3\,}
\left|\frac{\lambda_{\chi\chi h}}{\,(4M_\chi^2-M_h^2)v\,}
\left[\sum_{{f=\textrm{SM}}} \hspace{-2mm} N_{cf}^{}Q_f^2F_{1/2}^{}
(\tau_{\chi f}^{})+F_{1}^{}(\tau_{\chi W})\right]
+\hspace{-1mm} \frac{\lambda_{\chi\chi S}^{}}{4M_\chi^2-M_S^2}\frac{\yt_S^{}}{\,\sqrt{2}M_\T^{}}
\!\!\sum_{{f=\T,\T'}} \hspace{-2mm} N_{cf}^{}Q_{f}^2F_{1/2}^{}(\tau_{\chi f}^{})\right|^2 \!\!\!,
~~~~~~~~~~~
\\[2mm]
(\sigma_A^{} v )_{Z\gamma}^{} \hspace{-2mm} &\!=\!& \hspace{-2mm}
\frac{\,8\alpha^2M_\chi^2\,}{\pi^3}
\left|\frac{\lambda_{\chi\chi h}}{\,(4M_\chi^2-M_h^2)v\,}
\left[\sum_{{f=\textrm{SM}}} {N_{cf}^{}Q_f^{}(T_f^{3L}-2Q_f s_W^2)} B_{1/2}^{}
(\tau_{\chi f}^{},\eta_f^{})+B_{1}^{}(\tau_{\chi W}^{},\eta_W^{})\right]
\right. \nonumber
\\[-1mm]
&& \hspace{14mm}
\left.
+\frac{\lambda_{\chi\chi S}^{}}{\,4M_\chi^2-M_S^2\,}
\frac{\yt_S^{}}{\,\sqrt{2}M_\T^{}\,}
\!\!\sum_{{f=\T,\T'}} {N_{cf}^{}Q_f^{}(T_f^{3L}-2Q_f s_W^2)}
B_{1/2}^{}(\tau_{\chi f}^{},\eta_f^{})\right|^2 \!,
\end{eqnarray}
\eeqs
where $\,\alpha=1/128$,\, $\tau_{\chi f}^{}=M_f^2/M_\X^2$,\,
$\,\eta_f^{}=4M_f^2/M_Z^2$,\, and $\,N_{cf}^{}=3\,(1)\,$ corresponds
to the color factor of quarks (leptons).
The loop factors $\,F_{1,1/2}^{}(\tau)$\, and
$\,B_{1,1/2}^{}(\tau,\eta)$\, are defined in Appendix\,A.
The upper bound on the annihilation $\,\chi\chi\to\gamma X\,$
can be extracted from galactic center $\gamma$-ray line search, i.e.,
Fermi-LAT in low photon energy range\,\cite{FermiLATline}
and HESS in high energy range\,\cite{HESSline}.
Provided that the DM annihilations into $\,\gamma\gamma$\, and \,$\gamma Z\,$
are the only sources to generate gamma ray line, we may implement the constraint
on the quantity $\,2(\sigma_A^{} v )_{\gamma\gamma}^{}+(\sigma_A^{} v )_{\gamma Z}^{}$.\,
\gfig{fig:7} presents this quantity as a function of $\,\MX\,$
for Sample-A [plot-(a)] and Sample-B [plot-(b)] by red curves.
The upper bounds of Fermi-LAT and HESS are around
$\,10^{-27}\,\textrm{cm}^{3}\textrm{s}^{-1}$.\,
In each plot, the red curve is our theory prediction and
the black dot represents our prediction after imposing
the constraint of the observed DM relic density.
Plot-(a) shows that our Sample-A prediction is fully safe from the bound of Fermi-LAT.
For Sample-B, the bound from HESS in plot-(b)
is also not yet strong enough, but is quite close to our prediction.

The second type of gamma ray signal is the diffuse continuum spectrum from
secondary production of photons from primary DM annihilations,
$\,\chi\chi\to W^+W^-,\, ZZ,\, b\bar{b},\, \tau^+\tau^-,\, \mu^+\mu^-$.\,
The secondary photon is then initiated from the final state radiation or hadronization
with decays $\,\pi^0\to\gamma\gamma$\,.\,
The latest results come from the 4-years data of Fermi-LAT observation of
15 Milky Way dwarf spheroidal satellite galaxies \cite{FermiLAT2013}.
In the near future, the next generation experiments with better angular resolution
(such as CTA \cite{CTA}) will largely improve the sensitivity
over a wider mass range. We may extract the conservative constraint
by taking into account of the branching ratio for each detection channel.
In Sample-A and Sample-B, these final states arise from the light Higgs exchange.
We find that the predicted cross sections are far below the current upper bounds from
Fermi-LAT and HESS. But their predictions are within the reach of future experiments
via $\,W^+W^-$\, and \,$ZZ$\, channels.

Another way of DM indirect detection is to measure
the cosmic ray antiprotons, which could be produced from hadronization
of the primary products of DM annihilations.
Considering the uncertainty in modeling the antiproton propagation in galaxies,
Ref.\,\cite{antiproton} derived limits on the annihilation cross sections with
$\,W^+W^-$ and $\,b\bar{b}\,$ final states from AMS02 $p/\bar{p}$ ratio
measurement \cite{AMS02}. The antiproton constraints are only slightly stronger
than that from Fermi-LAT in the small mass region,
$\,\MX \lesssim 200\,$GeV.\, They impose no real constraints on our samples.
In passing, the antiproton constraint on $gg$ final state was
discussed in \cite{Hektor:2016uth}, showing that the $gg$ final state is dominant
at high energy end of the spectrum and AMS02 may have potential to probe this signature.

\section{Conclusions}
\label{sec:conclusion}
\label{sec:6}

The observed diphoton excess at the LHC Run-2 \cite{ATLAS}\cite{CMS},
if confirmed, would point to an exciting direction of new physics beyond the SM.
In this work, we constructed a minimal model which is well motivated by
realizing dark matter candidate, ensuring vacuum stability, and generating cosmic inflation.
With this we provided an explanation of the recently observed
$750\,$GeV new resonance at the LHC Run-2. In addition to the SM particle spectrum,
our model contains one complex singlet scalar $\,\S\,$ and one vector-like weak doublet
quark $\,\TT =(\T'\!,\,\T)^T$.\,
The real component $\,S\,$ of the singlet $\,\S\,$ has Yukawa interaction
with $\,\TT\,$ and can act as the $750\,$GeV resonance.
Since $\,(\T'\!,\,\T)$\, carry hypercharge $\,\fr{7}{6}\,$ and thus larger electric charges
$(\fr{5}{3},\,\fr{2}{3})$ than the SM quark doublet,\,
this makes $\,S\,$ have larger decay rate into diphotons.
We demonstrated that $\,S$\, can serve as the $750\,$GeV new resonance and explain
the observed excess of diphoton signals. Furthermore, the imaginary component $\,\chi\,$ of
the singlet $\,\S\,$ is a CP-odd pseudoscalar and the CP symmetry ensures $\,\chi\,$ to be
a stable DM candidate.
We find that this construction is rather economical and predictive, where the free parameters
in the scalar potential are almost fully determined by accommodating the $750$\,GeV resonance,
the vacuum stability of scalar potential, and the DM relic abundance. For explicit demonstration,
we constructed two numerical Samples A and B to study the phenomenology.
In section\,\ref{sec:3}, we analyzed the prediction of $\,S(750\text{GeV})\,$,\,
including its production and decays at the LHC Run-2. Further tests will be given
by the upcoming LHC runs in this year. Then, in section\,\ref{sec:4}
we studied the constraints of vacuum stability
and the realization of Higgs inflation in our model.
We derived the constraints from the DM relic abundance in section\,\ref{sec:5.1},
and analyzed the mono-jet signals
$\,pp\to j\,S\to j\,\X\X\,$ at the LHC Run-2 in section\,\ref{sec:5.2}.
Finally, we presented the constraints from the DM direct and indirect searches
in sections\,\ref{sec:5.3}--\ref{sec:5.4}.

\begin{appendix}
\section{Formulas for New Scalar Decays}
\label{app1}

In this Appendix, we derive the partial widths of
the heavy scalar $S$ decaying into the SM particles
(gauge bosons, Higgs bosons and fermions).
We present the following general formulas, which are used in our
current analyses,
\begin{subequations}
\begin{eqnarray}
  \Gamma(S \!\!\to\! hh)
& \!\!=\!\! &
  \frac {G^2_{Shh}}{32 \pi M_S}
  \sqrt{1 \!- \frac {4 M^2_h}{M^2_S}\,} \,,
\\
\label{eq:A1b}
  \Gamma (S\!\!\to\!f \bar f)
& \!\!=\!\! &
  \xi^2_{Sff}
  \frac{\,N_c g^2 m^2_f\,}{32 \pi M^2_W} M_S^{}
  \left( 1 - 4 \frac {m^2_f}{M^2_S} \right)^{3/2} ,
\\
  \Gamma (S\!\!\to\! WW)
& \!\!=\!\! &
  \xi^2_{SWW}
  \frac{g^2 M_S}{64 \pi} \frac{\sqrt{1 \!-\! x_W^{}}}{x_W^{}}
  \left( 4 - 4 x_W + 3 x^2_W \right) ,
\label{eq:width-SWW}
\\
  \Gamma (S \!\!\to\! ZZ)
& \!\!=\!\! &
  \xi^2_{S ZZ}
  \frac{g^2 M_S}{128 \pi} \frac {\sqrt{1 \!-\! x_Z^{}}}{x_W^{}}
  \left( 4 - 4 x_Z^{} + 3 x^2_Z \right) ,
\label{eq:width-SZZ}
\\
  \Gamma (S \!\!\to\! \gamma\gamma)
& \!\!=\!\! &
  \frac {\alpha^2 g^2}{256 \pi^3} \frac {M_S}{x_W}
\left| \sum_f N_{cj} Q^2_f \xi_{S f f}^{}
F_{\!1/2}^{}(\tau_f^{}) + \xi_{SWW}^{} F_1^{} (\tau_W) \right|^2 ,
\\
\label{eq:A1f}
  \Gamma (S\!\!\to\! gg)
& \!\!=\!\! &
  \frac{\alpha^2_s g^2}{128 \pi^3} \frac{M_S^{}}{x_W^{}}
\left| \sum_f \xi_{S f f}^{} F_{1/2}^{}(\tau_f^{}) \right|^2 ,
\\
  \Gamma (S \!\!\to\! Z\gamma)
& \!\!=\!\! &
  \frac{\alpha^2 g^2}{128 \pi^3} \frac {M_S}{x_W}
\left| \sum_f  \xi_{S f f}^{}N_{cf}^{}Q_f^{}(T_f^{3L}-2Q_f s_W^2) B_{1/2}^{}(\tau_f^{},\eta_f^{})
+ \xi_{SWW}^{} B_1^{}(\tau_W^{}, \eta_W^{}) \right|^2 ,
\end{eqnarray}
\label{eq:Gamma-scalar}
\label{eq:A1}
\end{subequations}
where $\,x_Z^{} = 4 M^2_Z / M^2_S$,\,
$x_W^{}= 4 M^2_W / M^2_S$,\, $\tau_{\!f}^{}=4M_f^2/M^2_S$,\,
$\,\eta_f^{}=4M_f^2/M_Z^2$,\,
and the color factor $\,N_{cf}^{}=3\,(1)\,$ for quarks (leptons).
For convenience, in the above we have rescaled the coupling ratios for $\,\T(\T')\,$
in Table\,\ref{tab:2} as,
$\,\xi_{S\T\T(S\T'\T')}^{}
 =c_\alpha^{}\yt_S^{}v/(\!\!\sqrt{2}M_{\T(\T')}^{})$.\,
The loop functions $\,F_{1}^{}(\tau_f^{})\,$ and $\,F_{1/2}^{}(\tau_f^{})\,$
are defined as,
\beqs
\begin{equation}
F_1^{} \,=\, 2 + 3 \tau \,[1 + (2\! - \tau) f(\tau)]\,,
\quad~~~
F_{\!1/2}^{} \,=\, - 2 \tau \,[1 + (1\! - \tau) f(\tau)]\,,
\end{equation}
with
\begin{equation}
  f(\tau) \,=\,
\begin{cases}
  ~\( \sin^{-1}\!\!\! \sqrt{1/\tau\,} \,\)^2 ,
  &~~ \mbox{if~\,} \tau \geqq 1 ,
  \\[1.5mm]
  ~- \frac 1 4 \left[ \ln \left( \eta_+^{}/\eta_-^{} \right) - i \pi \right]^2,
  &~~ \mbox{if~\,} \tau < 1 ,
\end{cases}
\end{equation}
\eeqs
where $\,\eta^{}_\pm = 1 \pm \!\sqrt{1 \!-\! \tau\,}$.\,
The loop functions $\,B_{1}^{}(\tau_{\!f}^{},\,\eta_f^{})\,$
and $\,B_{1/2}^{}(\tau_{\!f}^{},\,\eta_f^{})\,$ are defined as,
\beqs
\begin{eqnarray}
  B_{1}^{}(\tau,\eta ) &\!\!=\!\!& -t_W^{-1}\left[4(3-t_W^2)I_2^{}(\tau,\eta)
  +\left((1+2\tau)t_W^2-(5+2\tau)\right)I_1^{}(\tau,\eta )\right] ,
\\[2mm]
  B_{1/2}^{}(\tau,\eta) &\!\!=\!\!&
  \frac{-2}{s_W^{}c_W^{}}[I_1^{}(\tau,\eta)-I_2^{}(\tau,\eta)]\,,
\\[2mm]
  I_1^{}(\tau,\eta) &\!\!=\!\!& \frac{\tau\eta}{\,2(\tau\!-\!\eta)\,}
  +\frac{\tau^{2}\eta^{2}}{\,2(\tau\!-\!\eta)^2\,}\left[f(\tau)\!-\! f(\eta )\right]
  +\frac{\tau^{2}\eta}{(\tau\!-\!\eta)^2}\left[g(\tau)\!-\! g(\eta )\right],
  \hspace*{10mm}
\\[2mm]
  I_2^{}(\tau,\eta ) &\!\!=\!\!&
  -\frac{\tau\eta}{2(\tau\!-\!\eta)}\left[f(\tau)-f(\eta)\right],
\\[2mm]
  g(\tau) &\!\!=\!\!& \left\{\begin{array}{ll}
  \sqrt{\tau\!-1\,}\arcsin\sqrt{1/\tau}\,, &~~~~ \tau\geqq 1,
  \\[3mm]
  \frac{1}{2}\!\sqrt{1\!-\tau\,}\left[\log(\eta_+/\eta_-)-i\pi\right],
  &~~~~ \tau <1 ,
  \end{array}\right.
\end{eqnarray}
\eeqs
where we have used the abbreviations,
$\,(s_W^{},\,c_W^{}) = (\sin\theta_W^{},\,\cos\theta_W^{})$\,
and $\,t_W^{} = \tan\theta_W^{}$\,,\, with $\,\theta_W^{}\,$ denoting the weak mixing angle.
For $\,h\!\to\! gg$\, in the SM, the QCD corrections will introduce an enhancement factor
of $\,K(M_h^{}) \approx 1.5$\, for $\,M_h^{}=750$\,GeV \cite{Spira:1995rr}\cite{HRev}.
For the current case of $\,S(750\text{GeV})$,\, the decay width of $\,S\!\to\! gg$\,
is generated by $\T(\T')$ triangle-loops instead of the top loop. But, the QCD $K$-factor
is expected to be similar to the SM case.
So we use the SM $K$-factor as a reasonable estimate,  $\,K(M_S^{}) \approx 1.5$\,.

There is one complication for the $S$ decays into weak gauge boson pairs $WW$ and $ZZ$.\,
At tree level, $S$ could couple with $WW$ and $ZZ$ through mixing with the SM Higgs boson $h$.\,
The corresponding decay widths are given in \geqn{eq:width-SWW} and \geqn{eq:width-SZZ},
respectively. But, they should vanish when the mixing angle $\,\alpha\to 0\,$,\, or,
becomes negligible for $\,\alpha \lesssim 10^{-3}$.\,
In this case, the $\,\T(\T')$\, triangle-loop corrections become dominant.
Here we derive the decay width up to one-loop level,
\begin{eqnarray}
\Gamma (S \!\to\!\! WW) \,=\,
\frac{\,|\mathcal{M}(S \!\!\to\! WW)|^2\,}{16 \pi M_S} \,,
\qquad
\Gamma (S \!\to\!\! ZZ) \,=\,
\frac{\,|\mathcal{M}(S \!\!\to\! ZZ)|^2\,}{32 \pi M_S} \,,
\label{eq:width-SVV}
\end{eqnarray}
with the decay matrix elements parametrized as,
\begin{subequations}
\beqa
\label{eq:decay-M}
\mathcal{M} (S \!\!\to\! V_j^{}V_{j'}^{}) \,=\,
\left\{\text{i}c_0^{}g^{\mu\nu}\!
- c_1^{}\! \left[(k_1^{}\!\cdot k_2^{})g^{\mu\nu}\!-k_1^{\nu}k_2^{\mu}\right]\right\}
\ep_{\mu j}^{}(k_1^{})\ep_{\nu j'}^{}(k_2^{})\,,
\eeqa
where $\,V=W,Z$,\, and $\,j,j'=(+,-,0)\,$ denote the three polarizations of
weak gauge boson $V^\mu$.\, In the above decay amplitude, the coefficients
$(c_0^{},\,c_1^{})$ are given by the tree-level and triangle-loop contributions,
respectively. For the loop contribution $c_1^{}$,\, we compute the triangle-loop
by setting the final state $VV\,$ be massless, which is well justified
due to $\,M_S^2\gg M_V^2\,$.\, We summarize the results as follows,
\begin{eqnarray}
S \!\!\to\! WW\!:
&&
  c_0^{} = \sin \alpha\, \frac {2 M^2_W} v \,,
\qquad
  c_1^{} =  \frac{\alpha N_c}{\,2 s^2_w \pi v\,} A(\tau) \,,
\\
  S \!\!\to\! ZZ\!:
&&
  c_0^{} = \sin \alpha\, \frac {2 M^2_Z} v \,,
\qquad~
  c_1^{} = \frac{\alpha N_c}{\,c^2_w s^2_w \pi v\,}
  \left( \frac 1 2 - s^2_w + \frac {29} 9 s^4_w \right) A(\tau) \,.
\end{eqnarray}
\label{eq:width-SVV-c}
\end{subequations}
Since the two vector-like quarks $\,\mathcal T\,$ and $\,\mathcal T'\,$ have nearly degenerate masses,
we have $\,\tau = 4 M^2_{\T} / M^2_S \simeq 4 M^2_{\mathcal T'}/M^2_S$,\,
and $\,A(\tau) = - \frac{1}{2}F_{\!1/2}^{}(\tau)$\,.\,
We also note that for the $\,c_1^{}$\, related loop contributions, the longitudinal polarization
has negligible contributions to the decay amplitude \eqref{eq:decay-M}.
Since $\,M_S^{}\gg M_V^{}$,\, we can apply the equivalence theorem\,\cite{ET}
to replace the final state longitudinal component $V_L^a$ by the corresponding
would-be Goldstone boson $\,\pi^a$.\, But we find that the Goldstone amplitude
is nearly vanishing because the Yukawa couplings of $\,\pi^a$ with $\T(\T')$
are highly suppressed by the tiny mixing angles $\theta_{Rj}^{}$
according to Eqs.\,\eqref{eq:LY}-\eqref{eq:mass-mix} and $\theta_{Rj}^{}$ formula
below them. Thus, for the triangle-loop contributions it is a good approximation
to treat the final state $VV$ to be massless and ignore the longitudinal polarization.
As a consistent check, we find that including longitudinal polarization to the
final state $VV$ at one-loop could only affect the partial decay width by about
$(1\!-\!2)\%$
and has negligible effect.

\gfig{fig:BR} clearly shows that the effect of triangle-loop contributions dominate
\,Br($S\!\!\to\! WW,ZZ$)\, and the $WW/ZZ$ branching fraction curves become nearly flat
over the small $\,\alpha\,$ region of $\,\alpha\lesssim 2\!\times\!10^{-3}$,\,
where the tree-level contributions are negligible
due to the severe $\,\sin\!\alpha$\, suppression.

\section{One-Loop $\,\beta$\, Functions from New Couplings}
\label{app2}

In this Appendix, we present the additional terms in the one-loop $\beta$ functions
which are induced by the new couplings of our model.
We first consider the couplings \,$(\lambda_1^{},\,y_{ij}^{},\,g_s^{},\,g,\,g')$,\,
which also appear in the SM.
We analyze their $\beta$ functions and find the following new terms,
\begin{equation}
\label{betafunc1}
\begin{aligned}
& \Delta\beta_{\lambda_1^{}}^{}=\frac{1}{(4\pi)^2}
  \bigg(\frac{1}{2}\lambda_3^2 +\! \frac{1}{2}\lambda_6^2\bigg)\,,~
&& \Delta\beta_{y_{t}^{}}=0\,,~
&& \Delta\beta_{g_s}=\frac{2}{3(4\pi)^2}g_s^2,~
&& \Delta\beta_{g}=\frac{2}{3(4\pi)^2}g^2,~
&& \Delta\beta_{g'}=\frac{49}{9(4\pi)^2}g'^2.
\end{aligned}
\end{equation}
Then, we derive the one-loop $\,\beta\,$ functions for the new couplings of our model,
{\allowdisplaybreaks
\begin{subequations}
\label{betafunc2}
\begin{align}
\beta_{\lambda_2}^{} =&~~
\frac{1}{(4\pi)^2}\bigg(18\lambda_2^2+2\lambda_3^2+\frac{1}{2}\lambda_5^2
+24\lambda_2^{} \yt_S^2-12\yt_S^4\bigg)\,,
\\
\beta_{\lambda_3}^{}=&~~\frac{1}{(4\pi)^2}
\bigg[\lambda_3^{}\bigg(12\lambda_1^{}+6\lambda_2^{}+4\lambda_3^{}+6y_t^2+12\yt_S^2
-\frac{9}{2}g^2-\frac{3}{2}g'^2\bigg)+\lambda_5^{}\lambda_6^{}\bigg]\,,
\\
\beta_{\lambda_4}^{}=&~~\frac{1}{(4\pi)^2}
\bigg(18\lambda_4^2+\frac{1}{2}\lambda_5^2+2\lambda_6^2\bigg)\,,
\\
\beta_{\lambda_5}^{} =&~~\frac{1}{(4\pi)^2}
\bigg(6\lambda_2^{}\lambda_5^{}+6\lambda_4^{}\lambda_5^{}
+4\lambda_3^{}\lambda_6^{}+12\lambda_5^{}\yt_S^2\bigg)\,,
\\
\beta_{\lambda_6}^{} =&~~
\frac{1}{(4\pi)^2}\bigg[\lambda_6^{}\bigg(12\lambda_1^{}+6\lambda_4^{}+4\lambda_6^{}
+6y_t^2+6\yt_S^2-\frac{9}{2}g^2-\frac{3}{2}g'^2\bigg)+\lambda_3^{}\lambda_5^{}\bigg]\,,
\\
\beta_{\yt_S^{}}^{}=&~\frac{\yt_S^{}}{(4\pi)^2}\bigg(\frac{9}{2}\yt_S^2-8g_s^2
-\frac{9}{4}g^2-\frac{49}{6}g'^2\bigg)\,.
\end{align}
\end{subequations}}
For the Yukawa couplings in the above formulas,
we only need to keep the top quark Yukawa coupling $y_t^{}$ and the heavy quark $\,\TT\,$
Yuwaka coupling $\,\yt_S^{}\,$,\, as explained in the text.

\end{appendix}

\vspace*{5mm}
\noindent
{\bf Acknowledgments}:\\[1.5mm]
We thank Mingshui Chen, Xin Chen, and Weiming Yao for discussing
the ATLAS and CMS data concerning the diphoton excess.
This work was supported in part by National NSF of China
(under grants Nos.\,11275101 and 11135003) and
National Basic Research Program (under grant No.\,2010CB833000).


\end{document}